\documentclass[lettersize,journal]{IEEEtran}
\usepackage{amsmath,amsfonts}
\usepackage{array}
\usepackage[caption=false,font=normalsize,labelfont=sf,textfont=sf]{subfig}
\usepackage{textcomp}
\usepackage{stfloats}
\usepackage{url}
\usepackage{verbatim}
\usepackage{graphicx}
\usepackage{cite}
\usepackage{algorithmic}
\usepackage{color}
\usepackage{xcolor}
\usepackage{soul}
\usepackage{algorithm}
\usepackage{adjustbox}
\usepackage{multirow}
\usepackage{bbding}
\usepackage{hyperref}
\hypersetup{hidelinks=true}
\newcommand{\etal}{\textit{et al.}}

\begin{document}

\title{\huge ConnectomeDiffuser: Generative AI Enables Brain Network Construction from Diffusion Tensor Imaging}

\author{
Xuhang Chen,
Michael Kwok-Po Ng,
Kim-Fung Tsang,
Chi-Man Pun and
Shuqiang Wang
\thanks{This work was supported by National Natural Science Foundations of China under Grant 62172403, 12326614.}
\thanks{Xuhang Chen, Kim-Fung Tsang and Shuqiang Wang are with Shenzhen Institutes of Advanced Technology, Chinese Academy of Sciences, Shenzhen 518055, China (e-mail: xx.chen2@siat.ac.cn; kftsang@ieee.org; sq.wang@siat.ac.cn).}
\thanks{Xuhang Chen and Chi-Man Pun are with Department of Computer and Information Science, University of Macau, Macau, SAR, China (e-mail: yc17491@umac.mo; cmpun@umac.mo).}
\thanks{Michael Kwok-Po Ng is with Department of Mathematics, Hong Kong Baptist University, Hong Kong (email: michael-ng@hkbu.edu.hk).}
\thanks{Chi-Man Pun and Shuqiang Wang are the corresponding authors (e-mail: cmpun@umac.mo; sq.wang@siat.ac.cn).}

}

\maketitle

\begin{abstract}
Brain network analysis plays a crucial role in diagnosing and monitoring neurodegenerative disorders such as Alzheimer's disease (AD). 
Existing approaches for constructing structural brain networks from diffusion tensor imaging (DTI) often rely on specialized toolkits that suffer from inherent limitations: operator subjectivity, labor-intensive workflows, and restricted capacity to capture complex topological features and disease-specific biomarkers.
To overcome these challenges and advance computational neuroimaging instrumentation, ConnectomeDiffuser is proposed as a novel diffusion-based framework for automated end-to-end brain network construction from DTI.
The proposed model combines three key components: (1) a Template Network that extracts topological features from 3D DTI scans using Riemannian geometric principles, (2) a diffusion model that generates comprehensive brain networks with enhanced topological fidelity, and (3) a Graph Convolutional Network classifier that incorporates disease-specific markers to improve diagnostic accuracy.
ConnectomeDiffuser demonstrates superior performance by capturing a broader range of structural connectivity and pathology-related information, enabling more sensitive analysis of individual variations in brain networks. 
Experimental validation on datasets representing two distinct neurodegenerative conditions demonstrates significant performance improvements over other brain network methods. 
This work contributes to the advancement of instrumentation in the context of neurological disorders, providing clinicians and researchers with a robust, generalizable measurement framework that facilitates more accurate diagnosis, deeper mechanistic understanding, and improved therapeutic monitoring of neurodegenerative diseases such as AD.
\end{abstract}


\section{Introduction}
\IEEEPARstart{A}{dvances} in neuroimaging technologies have revolutionized the diagnosis and understanding of neurodegenerative diseases such as Alzheimer's disease (AD). The growing of medical imaging methods such as Functional Magnetic Resonance Imaging (fMRI) and other neuroimaging modalities has provided unprecedented insights into brain morphology and pathophysiology \cite{wang2018classification,wang2019ensemble,lei2022predicting}. Learning-based computational imaging has transformed multiple facets of medical image analysis, including lesion detection \cite{jit2,wang2017automatic}, tissue segmentation \cite{jit1}, image reconstruction \cite{hu20233,hu2020medical,li2022aivus,kelkar2021compressible}, super-resolution enhancement \cite{you2022fine,chen2022spatial,wang2025generative,you2022brain}, and diagnostic decision support \cite{wang2020ensemble,tce3,wang2025smart}. Similar transformer-based approaches have shown success in other imaging domains \cite{chen2024uwformer,guo2025underwater,dong2024multi}. Despite these advances, the extraction and interpretation of high-dimensional patterns within neuroimaging data remain challenges, particularly for neurodegenerative conditions characterized by subtle and progressive alterations in brain structure and function. Similar challenges exist in other medical domains, such as identifying biomarkers for osteoporosis through integrated genomic approaches \cite{zhu2021integrating}.

The human brain represents an intricate network of interconnected neurons, requiring precise coordination across anatomically distinct regions for optimal cognitive function. Conceptualizing the brain as a network structure has emerged as a transformative approach in neuroimaging \cite{tce4,tce5}, offering a more accurate representation of brain features. This network-based framework models the brain as a graph where vertices represent regions of interest (ROIs) and edges signify the interactions between these regions. The two primary forms of connectivity—functional connectivity (FC) and structural connectivity (SC)—have demonstrated exceptional sensitivity in detecting disease-specific alterations that can not be detected by conventional medical imaging \cite{jeon2020enriched,jing2023ta}. However, current methodologies for constructing brain networks frequently depend on specialized preprocessing toolkits such as PANDA \cite{cui2013panda}, which suffer from inherent limitations: procedural complexity, inconsistent outcomes due to parameter variability, and susceptibility to processing errors introduced by manual intervention. These shortcomings compromise the reliability of brain network analyses for neurological disorders and underscore the necessity for more robust, automated, and generalizable approaches.

To overcome these limitations, ConnectomeDiffuser is proposed as a comprehensive end-to-end framework that automates and refines the process of converting neuroimaging data into brain network representations. Designed specifically for brain disorder characterization, our framework integrates three components: a Template Network, a diffusion-based generative model, and a Graph Convolutional Network (GCN) classifier, collectively forming a robust pipeline for brain network construction and analysis.

The Template Network forms the foundation of our framework, extracting geometrically informed graph-structured features from three-dimensional Diffusion Tensor Imaging (DTI) data. This component employs the Anatomical Automatic Labeling (AAL) template \cite{desikan2006automated} to ensure anatomical precision while capturing complex non-linear fiber tract relationships. Drawing inspiration from Physics-Informed Neural Networks (PINNs) \cite{raissi2019physics}, the Template Network incorporates self-attention mechanisms to effectively model the intricate spatial relationships between neuronal pathways.

Building upon these extracted features, the diffusion-based model generates comprehensive brain networks that incorporate disease-specific characteristics. This generative component demonstrates superior robustness and fidelity in producing biologically plausible graph topologies compared to alternative generative approaches such as Generative Adversarial Networks (GANs) \cite{NIPS2014_5ca3e9b1,wang2024enhanced} and Variational Autoencoders (VAEs) \cite{vae}. The resulting connectivity matrices accurately represent the complex interrelationships between brain regions while preserving pathologically relevant patterns.

The GCN classifier completes our framework, analyzing the generated graph representations to capture both local and global connectivity patterns. This component's adaptability to diverse topological structures enables accurate categorization of different brain disorders based on their brain network abnormalities. The integration of the Template Network with the GCN classifier provides an unprecedented detailed characterization of brain connectivity, enhancing diagnostic accuracy for neurological disorders such as AD.

Our main contributions are as follows:
\begin{enumerate}
\item The ConnectomeDiffuser is proposed as an end-to-end method to identify the biomarkers for neurological disorders by transforming neuroimaging data into graph-based brain connectivity representations.
\item The Template Network is proposed to generate anatomically informed brain feature matrices using Riemannian geometric principles.
\item The Brain Diffusive Loss is proposed to precisely quantify and minimize topological discrepancies in the generated networks, significantly enhancing the model's ability to identify the complex connectivity patterns characteristic of various neurodegenerative diseases.
\end{enumerate}

\section{Related Work}
\subsection{Brain Network Analysis}
Graph Neural Networks (GNNs) have emerged as powerful computational tools for modeling brain networks, primarily due to their inherent capacity to capture complex, high-order interactions among regions of interest (ROIs). Among various GNN architectures, Graph Convolutional Networks (GCNs) have gained particular prominence in neuroimaging analysis. Parisot \etal \cite{parisot2018disease} and Ktena \etal \cite{ktena2017distance} demonstrated GCNs' effectiveness in detecting aberrant neuronal connections associated with Alzheimer's disease (AD), establishing their utility for neurological disorder characterization. Zuo \etal \cite{zuo2024bdht} leveraged generative AI approaches to enable causality analysis for mild cognitive impairment, further advancing the application of AI in understanding neurodegenerative disease progression.

Recent advances have further extended GNN applications in brain network modeling. Pan \etal \cite{pan2021decgan} introduced the Decoupling Generative Adversarial Network (DecGAN), which effectively identifies abnormal neural circuits implicated in AD pathophysiology. Their subsequent work on CT-GAN \cite{pan2022cross} leveraged cross-modal Transformer architectures to integrate functional connectivity from resting-state fMRI with structural connectivity derived from DTI, enhancing the comprehensiveness of brain network representations. Kong \etal \cite{kong2022adversarial} developed the Structural Brain-Network Generative Model (SBGM), employing an adversarial learning framework to generate structural brain networks directly from neuroimaging data, reducing reliance on conventional processing pipelines. Gong \etal \cite{gong2023addictive} introduced spatial attention recurrent networks with feature selection for identifying addiction-related brain networks, demonstrating the versatility of neural network approaches in characterizing various neurological conditions. More recently, Jing \etal \cite{jing2024addiction} proposed Graph Diffusion Reconstruction Network for identifying addiction-related brain networks, highlighting the emerging integration of diffusion techniques with graph-based brain network analysis.

Despite these significant advancements, current GCN-based methodologies face persistent challenges in fully capturing the intricate topological characteristics of brain networks. Many approaches inadequately incorporate disease-specific information and biomarkers, limiting their translational potential in computational neuroimaging for clinical diagnosis of neurological disorders. Our work addresses these limitations by adopting GCN as a specialized classifier for extracting disease-related embeddings while enhancing its capabilities through strategic integration with a Template Network and a diffusion-based generative model.

\subsection{Diffusion Models}
Diffusion models \cite{sohl2015deep} represent a class of probabilistic generative frameworks designed to learn data distributions by iteratively reducing noise in normally distributed data. These models have demonstrated exceptional promise across diverse medical imaging applications. Wolleb \etal \cite{Wolleb2022DiffusionMF} pioneered a weakly supervised approach for anomaly detection using Denoising Diffusion Implicit Models, generating highly detailed anomaly maps without requiring complex training regimens.

In neuroimaging specifically, Pinaya \etal \cite{Chung2021ScorebasedDM} established the clinical viability of diffusion models, while Chung \etal \cite{jo2022score} developed a sophisticated score-based diffusion model for three-dimensional medical image reconstruction. Their approach integrates denoising score matching to train time-dependent score functions with an iterative inference process that alternates between numerical Stochastic Differential Equation (SDE) solvers and data consistency steps. Gonzalez-Jimenez \etal \cite{gonzalez2023sano} extended these concepts to dermatological applications, employing an unsupervised diffusion model to generate log-likelihood gradient maps that highlight pathological regions indicative of eczema.

The synthesis of medical images represents another domain where diffusion models have demonstrated considerable utility. Han \etal \cite{han2023contrastive} developed a coarse-to-fine Positron Emission Tomography (PET) reconstruction framework incorporating diffusion techniques, achieving superior sample quality and higher log-likelihood scores compared to conventional GAN-based approaches. Yao \etal \cite{yao2023conditional} specifically applied conditional diffusion models for Alzheimer's disease prediction through data augmentation, demonstrating the potential of diffusion-based approaches in neurodegenerative disease analysis. The Latent Diffusion Model (LDM) \cite{rombach2022high}, an advanced generative architecture capable of capturing complex statistical characteristics and data structures, has found applications in clinical imaging. Wang \etal \cite{wang2023inversesr} exemplified this through InverseSR, an LDM-based methodology for enhancing the resolution of low-resolution functional MRI scans.

While diffusion-based approaches have demonstrated considerable promise across various medical imaging tasks, their application to the generalized construction of brain networks for neurological disorders remains largely unexplored. Current methodologies frequently lack the capacity to effectively incorporate disease-specific information and often inadequately capture the complex topological characteristics inherent to brain networks. Our work leverages the superior generative capabilities of diffusion models specifically for brain network construction, aiming to provide robust stability while addressing the limitations of existing approaches. Through the strategic integration of a Template Network and a GCN classifier, our proposed model enables the generalized construction of brain networks with enhanced fidelity, ultimately contributing to improved diagnosis and understanding of neurological disorders.

\begin{figure*}[ht]
    \begin{minipage}[b]{1.0\linewidth}
        \includegraphics[width=\linewidth]{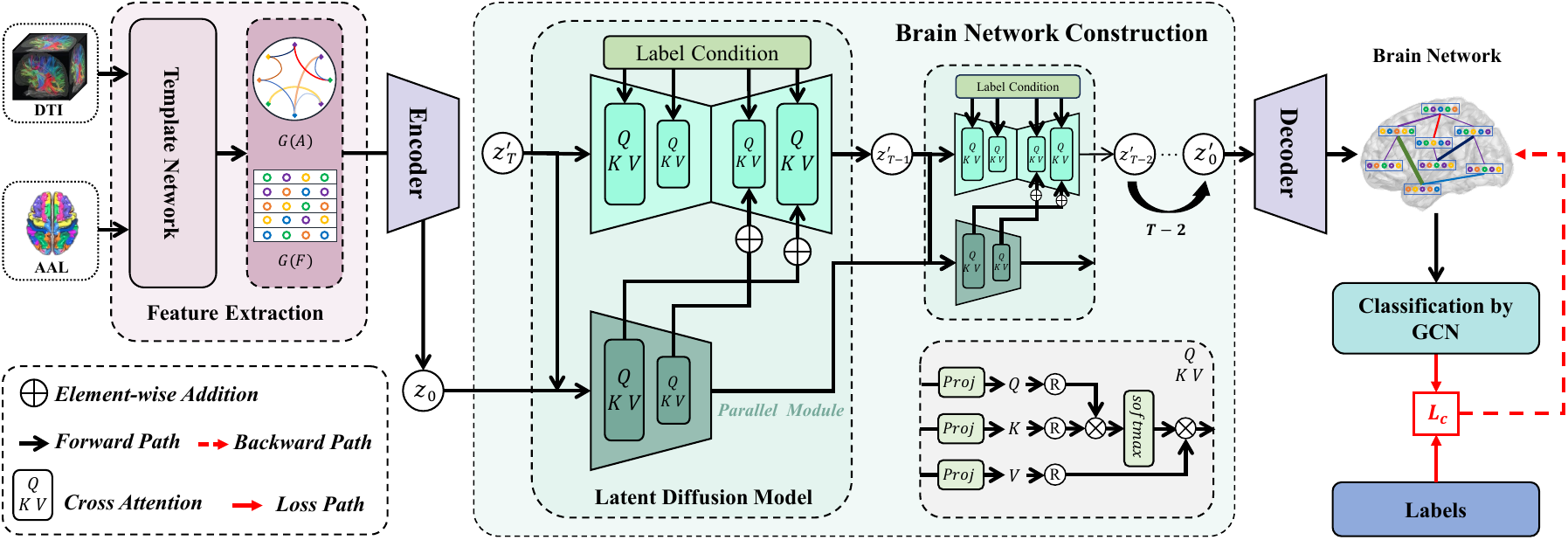}
    \end{minipage}
    \caption{
    The overall architecture of the proposed ConnectcomeDiffuser comprises several steps. Initially, the Template Network is applied to DTI images to extract topological features of the brain network. These features are then passed to the diffusion model, resulting in the construction of a comprehensive brain network. Subsequently, the GCN classifier is employed, incorporating classification knowledge to further enhance the feature extraction and diffusion model.
    }
    \label{fig:model}
\end{figure*}

\section{Method}

\subsection{Overview}
The proposed ConnectomeDiffuser framework comprises three principal components: the Template Network, a Latent Diffusion model, and a Graph Convolutional Network (GCN) classifier. As illustrated in Figure \ref{fig:model}, the processing pipeline begins with the registration of Diffusion Tensor Imaging (DTI) data to the Anatomical Automatic Labeling (AAL) template, followed by feature extraction via the Template Network. Subsequently, these extracted features are processed through the ConnectomeDiffuser to generate a comprehensive brain network representation. Finally, the GCN classifier analyzes the vertex structure of the brain features to perform diagnostic categorization.

\subsection{Anatomical Automatic Labeling Template}
The input to our model consists of Diffusion Tensor Imaging (DTI) data, denoted as $x \in \mathbb{R}^{h \times w \times d}$, representing a multidimensional array of diffusion-weighted measurements that characterize water molecule diffusion properties within cerebral tissue. Our objective is to transform this input into a brain network representation $y \in \text{Graph}_{E,V}$, where $E$ and $V$ denote the edges and vertices of the graph, respectively. Essentially, we aim to establish a complex non-linear mapping function $f: x \rightarrow y$ that translates neuroimaging data into a topologically meaningful graph structure.

To facilitate precise anatomical localization within the DTI data, we integrate the AAL template, which partitions the brain into 90 distinct neuroanatomical regions. This standardized parcellation scheme provides a consistent reference framework, enabling systematic alignment of DTI measurements with specific brain structures across subjects.

The registration process encompasses three sequential operations: (1) spatial normalization of DTI data to the template coordinate space, (2) identification of brain regions according to the AAL template labels, and (3) extraction of region-specific DTI information through element-wise multiplication between the template mask and the DTI data. This procedure yields an initial anatomical matrix $F \in \mathbb{R}^{90 \times \text{feature\_dim}}$, where $\text{feature\_dim}=128$, encapsulating a high-dimensional feature representation for each of the 90 brain regions.

The region-specific DTI features extracted through this registration process serve as the foundation for subsequent processing by the Template Network, which further refines and transforms these features into a more informative representation of brain connectivity patterns.

\subsection{Template Network}
The Template Network is designed to extract meaningful graph-structured features from the region-specific DTI data. While conventional approaches might employ standard convolutional neural networks, we implement a geometric approach that more effectively captures the complex fiber structures in the brain.

The brain's white matter pathways can be effectively modeled as geodesics (shortest paths) on a Riemannian manifold, which is a mathematical space where distances are measured along curved paths \cite{bihonegn2021geodesic,bihonegn20204th,o2002new}. This approach is particularly suitable for DTI data, as it naturally represents the directional diffusion of water molecules along fiber tracts.

\begin{figure}[ht]
    \begin{minipage}[b]{1.0\linewidth}
        \centering
        \includegraphics[width=\linewidth]
        {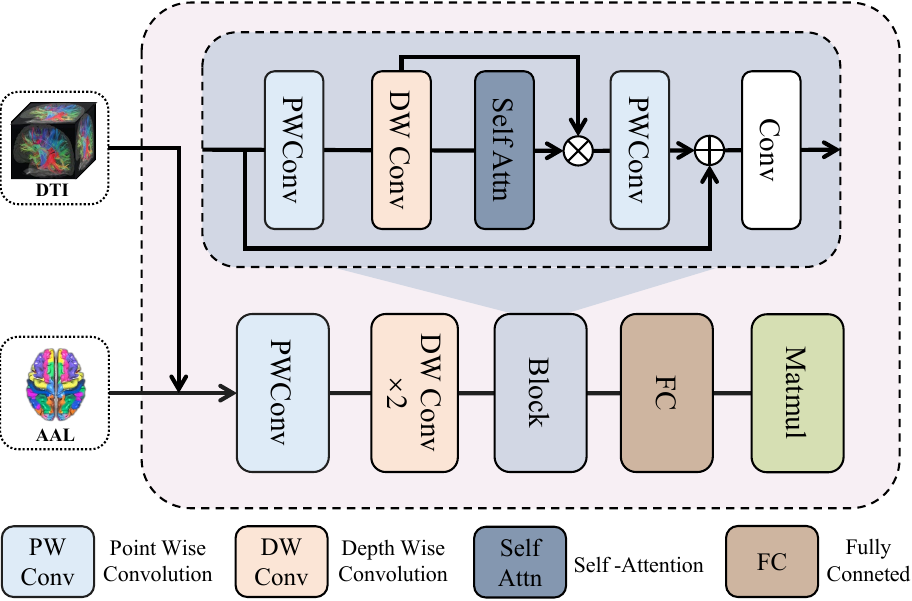}
    \end{minipage}
    \centering
    \caption{
    The architecture of the Template Network involves processing a tensor format of a 3D DTI image and an AAL template as input to generate the brain feature matrix as output.
    }
    \label{fig:FEN}
\end{figure}

As depicted in Figure \ref{fig:FEN}, the Template Network processes the registered DTI data through a series of specialized convolutional blocks and self-attention mechanisms. The network architecture consists of pointwise and depthwise convolutional layers for local feature processing, self-attention blocks to capture long-range dependencies, and fully connected layers to generate the final brain feature matrix.

The network transforms the initial anatomical matrix into two key outputs: (1) a brain feature matrix $F \in \mathbb{R}^{90 \times 80}$, where each of the 90 brain regions is represented by an 80-dimensional feature vector, and (2) an initial adjacency matrix $A \in \mathbb{R}^{90 \times 90}$ representing preliminary connectivity between brain regions.

To preserve the geometric properties of brain fiber structures, we model the output as a Riemannian metric, which is a symmetric positive-definite (SPD) matrix. This matrix encodes distances on the manifold and is constructed using eigenvalue decomposition:
\begin{equation}
g=\mathbf{R}\mathbf {\Lambda}\mathbf{R}^T,
\end{equation}
where $\mathbf{R}$ is a rotation matrix that aligns with the principal diffusion directions, and $\Lambda$ is a diagonal matrix with positive entries representing diffusion magnitudes. The parameters of $\mathbf{R}$ and $\Lambda$ are learned from the data.

The network is optimized to minimize the discrepancy between predicted geodesics (fiber pathways) and actual fiber orientations observed in the DTI data, effectively learning the underlying geometric structure of the brain's white matter.

\subsection{ConnectomeDiffuser}
Diffusion Models \cite{sohl2015deep} are probabilistic generative models that learn data distributions by iteratively denoising normally distributed variables through a fixed-length Markov Chain of length $T$. For image generation, certain diffusion models \cite{dhariwal2021diffusion,ho2020denoising} employ a reweighted variant of the variational lower bound, inspired by denoising score-matching models \cite{song2020score}, to capture complex patterns and generate high-quality outputs.

\begin{algorithm*}[!ht]
\caption{The optimization procedure for the ConnectomeDiffuser}
\label{alg1}
\begin{algorithmic}[1]

\REQUIRE DTI data for each subject: $\text{DTI}_i$ for $i = 1$ to $M$ ($M$ = number of subjects); AAL template defining $V$ brain regions ($V = 90$); Empirical brain connections: $\hat{A}_i$ for $i = 1$ to $M$; Number of GCN layers: $K$; Batch size: $BS$; Maximum number of epochs: $\text{Epoch}$; Hyperparameters: $\alpha$ and $\beta$.

\ENSURE Generated brain networks: $G(A, F)$; Predicted disease labels: $l$

\FOR{$s = 1$ to $\text{Epoch}$}
\STATE Sample batch of $B$ subjects $\{i_1, i_2, \ldots, i_B\}$

\FOR{$b = 1$ to $B$}
\STATE \textbf{// Step 1: AAL Template Registration}
\STATE Initialize $F_b = \text{zeros}(V, \text{feature\_dim})$
\FOR{$v = 1$ to $V$}
\STATE $F_b[v, :] = \text{DTI}_{i_b} \odot \text{AAL}[v]$ \COMMENT{Element-wise multiplication}
\ENDFOR

\STATE \textbf{// Step 2: Brain Feature Extraction via Template Network}
\STATE $F_b^1 = \text{Conv}(F_b)$
\STATE $F_b^2 = \text{Conv}(\text{Conv}(F_b^1))$

\FOR{$j = 1$ to $3$}
\STATE $a_b^j = \text{Sigmoid}(\text{Conv}(F_b^2))$
\STATE $F_b^2 = a_b^j \odot F_b^2$ \COMMENT{Apply attention weights}
\ENDFOR

\STATE $A_b^1 = \text{Linear}(F_b^2)$ \COMMENT{Project to adjacency space}
\STATE $A_b^2 = \text{Dropout}(A_b^1)$ \COMMENT{Apply dropout for regularization}
\STATE $A_b^3 = \text{MatMul}(A_b^2, \text{Transpose}(A_b^2))$ \COMMENT{Ensure symmetry}

\STATE $F_{\text{out},b} = F_b^2$ \COMMENT{Brain feature matrix ($V \times \text{feature\_dim}$)}
\STATE $A_{\text{init},b} = A_b^3$ \COMMENT{Initial adjacency matrix ($V \times V$)}

\STATE \textbf{// Step 3: Measure Riemannian Metric based on Geodesics}
\STATE Compute $L_{FE} = \sum_{i=1}^{m} ||\nabla^g_{v_i} v_i - \sigma_i v_i||^2 + \alpha \text{Reg}(g)$

\STATE \textbf{// Step 4: Diffusion process}
\STATE $z_{0,b} = E(F_{\text{out},b})$ \COMMENT{Encode features to latent space}

\STATE Sample $t \sim \text{Uniform}(1, T)$
\STATE Sample $\epsilon \sim \mathcal{N}(0, I)$
\STATE $z_{t,b} = \sqrt{\bar{\alpha}_t}z_{0,b} + \sqrt{1-\bar{\alpha}_t}\epsilon$ \COMMENT{Forward diffusion}

\STATE $\epsilon_{\text{pred}} = \epsilon_{\theta}(z_{t,b}, t, \tau_{\theta}(y_b))$ \COMMENT{Predict noise}

\STATE \textbf{// Step 5: Generate brain network}
\STATE $G_b = D(z_{0,b})$ \COMMENT{$D$ is the decoder}

\STATE \textbf{// Step 6: Predict disease category using GCN}
\STATE $\tilde{A}_b = \tilde{D}^{-1/2}(A_{\text{init},b} + I)\tilde{D}^{-1/2}$ \COMMENT{Laplacian transformation}

\STATE Initialize $H_b^0 = F_{\text{out},b}$
\FOR{$k = 1$ to $K$}
\STATE $H_b^k = \text{Mish}(\tilde{A}_b \cdot \text{Mish}(\tilde{A}_b \cdot H_b^{k-1} \cdot W_1^k) \cdot W_2^k)$
\STATE $h_b^k = \frac{1}{V}\sum_{j=1}^{V} H_b^k[j, :]$ \COMMENT{Global average pooling}
\ENDFOR

\STATE $h_b = \text{concat}(h_b^1, h_b^2, \ldots, h_b^K)$
\STATE $l_b = \text{FC}(h_b)$ \COMMENT{Predicted label}
\ENDFOR

\STATE \textbf{// Calculate losses}
\STATE $L_{\text{total}} = L_{LDM} + L_{FE} + L_C$

\ENDFOR

\RETURN Generated brain networks $G$ and predicted labels $l$
\end{algorithmic}
\end{algorithm*}

The diffusion model comprises an encoder $E$ and a decoder $\mathcal{D}$ that focus on the semantic content of the data while optimizing computational resources. The brain feature matrix $P \in \mathbb{R}^{90 \times 80}$ is encoded by $E$ into a latent representation $z = E(P)$. The decoder $D$ subsequently reconstructs the brain network from this latent representation, yielding $\tilde{P} = D(z) = D(E(P))$, where $z \in \mathbb{R}^{h \times w}$. The encoder downsamples the input by a factor $f = H/h = W/w$, with $f = 2^m, m \in \mathbb{N}$. The latent representation is processed by the diffusion module, and the decoder transforms the output into a brain network $B \in \mathbb{R}^{90 \times 90}$.

In the diffusion process, Gaussian noise with variance $\beta_t \in (0, 1)$ is added to the encoded latent representation $z = E(x)$ at time $t$ to produce the noisy latent representation:
\begin{equation}
    z_t=\sqrt{\bar{\alpha}_t}z+\sqrt{1-\bar{\alpha}_t}\epsilon,
\end{equation}
where $\bar{\alpha}_t=\prod_{s=1}^t\alpha_s$, $\epsilon\sim\mathcal{N}(0,\mathbf{I})$ and $\alpha_t=1-\beta_t$. As $t$ increases, $z_t$ approximates a standard Gaussian distribution.

During the diffusion process, we employ a U-Net \cite{ronneberger2015u} denoiser comprising an encoder, a middle block, and a decoder. The noisy latent variable is combined with the encoder output and processed through the model to obtain a clean latent representation $z_0^{\prime}$:
\begin{equation}
    \begin{aligned}
    z_0^{\prime}=\frac{z_t}{\sqrt{\bar{\alpha}_t}}-\frac{\sqrt{1-\bar{\alpha}_t}\epsilon_\theta(z_t,c,t,E(F))}{\sqrt{\bar{\alpha}_t}},
    \end{aligned}
\end{equation}
where $c$ represents conditioning information and $F$ denotes the brain feature matrix.

To enhance adaptability, the U-Net is combined with cross-attention mechanisms \cite{vaswani2017attention}. A domain-specific encoder $\tau_\theta$ projects the input to an intermediate representation $\tau_\theta(y) \in \mathbb{R}^{M \times d_\tau}$, which is integrated into the U-Net's intermediate layers via cross-attention:
\begin{equation}
Q = W^{(i)}_Q \cdot  \varphi_i(z_t), \; K = W^{(i)}_K \cdot \tau_\theta(y),
  \; V = W^{(i)}_V \cdot \tau_\theta(y),
\label{eqn:qkv}
\end{equation}
where $\varphi_i(z_t) \in \mathbb{R}^{N \times d^i_\epsilon}$ represents a flattened intermediate representation of the U-Net implementing $\epsilon_\theta$, and $W^{(i)}_V \in \mathbb{R}^{d \times d^i_\epsilon}$, $W^{(i)}_Q \in \mathbb{R}^{d \times d_\tau}$, and $W^{(i)}_K \in \mathbb{R}^{d \times d_\tau}$ are learnable projection matrices \cite{jaegle2021perceiver}.

We further enhance the diffusion model with a parallel module that shares the same architecture as the U-Net's middle block. This module processes disease-specific conditional information concurrently with the latent representation, enhancing the model's ability to generate condition-consistent brain networks. The outputs of this parallel module are integrated into the U-Net through adaptive convolutional layers at each scale.

During inference, the model samples $z_T$ from a standard normal distribution and iteratively applies the reverse diffusion process to obtain $z_0$. The decoder then transforms this latent representation into the final brain network $G = D(z_0)$.

\subsection{GCN Classifier}
The proposed GCN classifier integrates a Graph Convolutional Network layer \cite{kipf2017semi} with a fully connected layer. The GCN layer aggregates structural characteristics from each vertex's neighborhood, enabling the model to capture both local and global connectivity patterns within the brain network.

The GCN follows a layer-wise propagation rule:
\begin{equation}
    H^{(l+1)}=\sigma\left(\tilde{D}^{-\frac{1}{2}} \tilde{A} \tilde{D}^{-\frac{1}{2}} H^{(l)} W^{(l)}\right),
\end{equation}
where $\tilde{A} = A + I_N$ represents the adjacency matrix with added self-connections, $I_N$ denotes the identity matrix, $W^{(l)}$ is a layer-specific trainable weight matrix, and $\tilde{D}_{i i} = \sum_j \tilde{A}_{ij}$ is the degree matrix. The activation function $\sigma(\cdot)$ is the ReLU function, defined as $\text{ReLU}(\cdot) = \max(0, \cdot)$. The matrix $H^{(l)} \in \mathbb{R}^{N \times D}$ represents the activations in the $l^{th}$ layer, with $H^{(0)} = X$ being the initial input features.

The fully connected layer classifies the entire graph based on the features extracted by the GCN layer. While the classifier's primary function is categorization, it also serves to guide the generation process by providing classification signals that inform the diffusion model.

By leveraging the GCN architecture, our model effectively captures the complex relationships between brain regions and their connectivity patterns, facilitating comprehensive analysis of structural alterations associated with various neurological conditions.

\subsection{Brain Diffusive Loss}
The Brain Diffusive Loss ($L_B$) is a composite loss function that guides the learning process by incorporating information from multiple components of our model. It consists of three essential components:

\textbf{1. Feature Extraction Loss ($L_{FE}$):}
This component ensures accurate registration and feature extraction from the DTI data:
\begin{equation}
	\label{eqn:pinn}
	L_{FE}=\sum_{i=1}^{m} \|\nabla^g_{\mathbf{v}_i} \mathbf{v}_i-\sigma_i\mathbf{v}_i\|_{2}+\alpha \operatorname{Reg}(g),
\end{equation}
where $g$ represents the Riemannian metric learned by the Template Network, $\mathbf{v}_i$ are vector fields corresponding to fiber directions, $\nabla^g_{\mathbf{v}_i} \mathbf{v}_i$ is the covariant derivative, measuring how vector fields change along geodesics,  $sigma_i$ is a scaling factor, $\operatorname{Reg}(g)$ is a regularization term that ensures smoothness, $\alpha$ is a weighting parameter.

\textbf{2. Latent Diffusion Loss ($L_{LDM}$):}
This component guides the diffusion model to generate realistic brain networks:
\begin{equation}
L_{LDM} = E_{E(x), y, \epsilon \sim \mathcal{N}(0, 1), t }\Big[ \Vert \epsilon - \epsilon_\theta(z_{t},t, \tau_\theta(y)) \Vert_{2}^{2}\Big] \, ,
\label{eq:learn}
\end{equation}
where $E$ represents the encoded DTI features, $y$ represents the disease labels, $\epsilon$ is a noise vector sampled from a standard Gaussian, $t$ is a time step in the diffusion process, $z_t$ is the noisy latent variable at time $t$, $\epsilon_\theta$ and $\tau_\theta$ are learnable functions.

\textbf{3. Classification Loss ($L_C$):}
This component ensures the generated brain networks contain disease-relevant information:
\begin{equation}
L_C = - \frac{1}{N} \sum_{i=1}^N{p(y_i|x_i) log[q(\hat{y}_i|x_i)]},
\label{eqn:gcn}
\end{equation}
where $N$ is the number of samples, $p(y_i|x_i)$ is the true label distribution, $q(\hat{y}_i|x_i)$ is the predicted label distribution.

The total Brain Diffusive Loss is the sum of these three components:
\begin{equation}
	L_B = L_{FE} + L_{LDM} + L_C
\end{equation}

By simultaneously optimizing these three aspects, our model learns to generate anatomically accurate, realistic, and diagnostically informative brain networks that effectively capture the complex structural characteristics associated with various neurological disorders.

\section{Experiment}

\subsection{Data Preparation and Preprocessing}
We evaluated the proposed ConnectomeDiffuser framework on two well-established neuroimaging datasets: the Alzheimer's Disease Neuroimaging Initiative (ADNI) \cite{weiner2017alzheimer} and the Autism Brain Imaging Data Exchange (ABIDE) \cite{di2014autism}. These datasets represent distinct neurodegenerative and neurodevelopmental conditions, allowing us to assess the generalizability of our approach.

The ADNI dataset (Table \ref{tab:ds1}) comprises 300 subjects stratified into four clinical categories: normal controls (NC, $n=80$), early mild cognitive impairment (EMCI, $n=80$), late mild cognitive impairment (LMCI, $n=76$), and Alzheimer's disease (AD, $n=64$). This stratification enables the analysis of progressive neurodegeneration across the AD continuum. The ABIDE dataset (Table \ref{tab:ds2}) includes 160 participants, evenly distributed between individuals with autism spectrum disorder (ASD, $n=80$) and matched neurotypical controls (NC, $n=80$).

Prior to analysis, all diffusion tensor imaging (DTI) data underwent rigorous preprocessing using the Pipeline for Analyzing braiN Diffusion imAges (PANDA) toolkit \cite{cui2013panda}. The preprocessing pipeline encompassed: (1) skull stripping to eliminate non-brain tissue, (2) motion correction to compensate for subject movement, (3) eddy current correction to mitigate gradient-induced distortions, and (4) tensor model fitting using a least-squares approach to compute diffusion metrics. This standardized preprocessing workflow ensured data consistency while minimizing artifacts that could potentially confound subsequent analyses.

We employed a stratified data partitioning strategy to maintain class distribution across training and testing sets. For the ADNI dataset, we allocated 220 subjects for model training and reserved 80 subjects (20 from each group) for testing. Similarly, the ABIDE dataset was divided into 148 subjects for training and 16 subjects (8 from each group) for testing. To enhance statistical reliability, all experiments were conducted using five-fold cross-validation, thereby mitigating potential biases from any particular data split.

\begin{table}[ht]
\caption{Detailed information of the ADNI dataset.}
    \centering
    \begin{tabular}{c c c c c}
    \hline
        Group & NC & EMCI & LMCI & AD \\ \hline
        Number & 80 & 80 & 76 & 64\\
        Gender & 38M/42F  & 41M/39F  & 36M/40F  & 31M/33F \\ 
        Age & 74.1 $\pm$ 6.4 & 75.4 $\pm$ 7.1 & 75.1 $\pm$ 4.6 & 76.2 $\pm$ 5.6 \\ \hline
    \end{tabular}
    \label{tab:ds1}
\end{table}

\begin{table}[ht]
\caption{Detailed information of the ABIDE dataset.}
    \centering
    \begin{tabular}{c c c}
    \hline
        Group & NC & ASD \\ \hline
        Number & 80 & 80 \\
        Gender & 65M/15F & 70M/10F \\ 
        Age & 13.2 $\pm$ 3.2 & 12.9 $\pm$ 2.7 \\ \hline
    \end{tabular}
    \label{tab:ds2}
\end{table}

\subsection{Implementation Details}
We implemented the ConnectomeDiffuser framework using PyTorch and trained the model on dual NVIDIA V100 GPUs. The optimization process employed the AdamW optimizer with default parameters, a learning rate of $1e-4$, and a batch size of 4. All models were trained for 300 epochs to ensure convergence.

The diffusion model component utilized a downsampling factor of 4 in the encoder and 100 diffusion steps, striking an optimal balance between computational efficiency and network generation quality. The U-Net architecture for denoising comprised a symmetric encoder-decoder structure with four resolution levels. Each level in the downsampling path contained two convolutional layers followed by downsampling, with the number of filters progressively increasing from 64 to 512 (specifically, 64, 128, 256, and 512 filters). The middle block incorporated two convolutional layers with 512 filters augmented by self-attention mechanisms to capture global context. The upsampling path mirrored this architecture in reverse, replacing downsampling operations with upsampling layers. All convolutional layers employed a kernel size of 3×3, instance normalization for stable training, and SiLU (Sigmoid Linear Unit) activation functions to introduce non-linearity.

\subsection{Evaluation Metrics}
To comprehensively assess the efficacy of ConnectomeDiffuser in brain network construction and neurological disorder classification, we employed four classification metrics:
\begin{enumerate}
    \item \textbf{Accuracy (ACC)}: This metric provides the proportion of true results among the total number of cases examined.
    \item \textbf{Precision (PRE)}: This metric measures the proportion of true positive identifications among all positive identifications made by the model.
    \item \textbf{Recall (REC)}: This metric assesses the proportion of actual positives correctly identified by the model.
    \item \textbf{F1 Score (F1)}: This metric is the harmonic mean of precision and recall.
\end{enumerate}
These metrics collectively provide a multifaceted evaluation of classification performance, enabling a robust comparison between ConnectomeDiffuser and alternative approaches.

\subsection{Classification Performance}
\begin{figure*}[ht]
    \begin{minipage}[b]{0.32\linewidth}
        \centering
        \includegraphics[width=\linewidth]{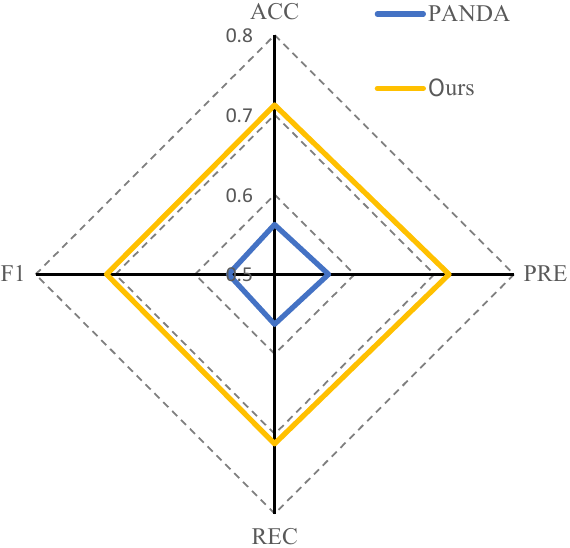}
        \medskip
    \end{minipage}
    \begin{minipage}[b]{0.32\linewidth}
        \centering
        \includegraphics[width=\linewidth]{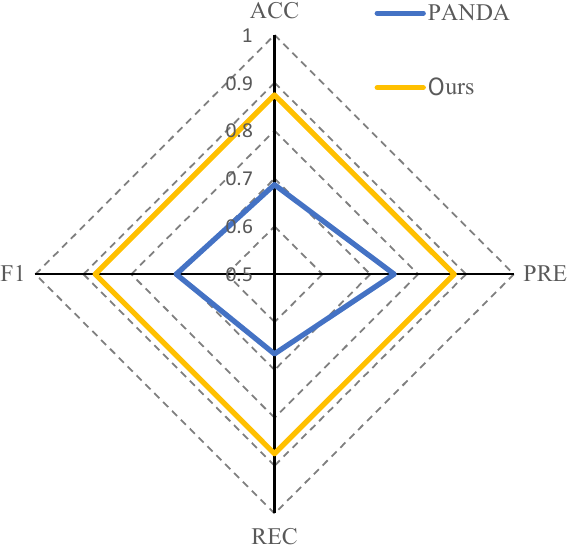}
        \medskip
    \end{minipage}
    \begin{minipage}[b]{0.32\linewidth}
        \centering
        \includegraphics[width=\linewidth]{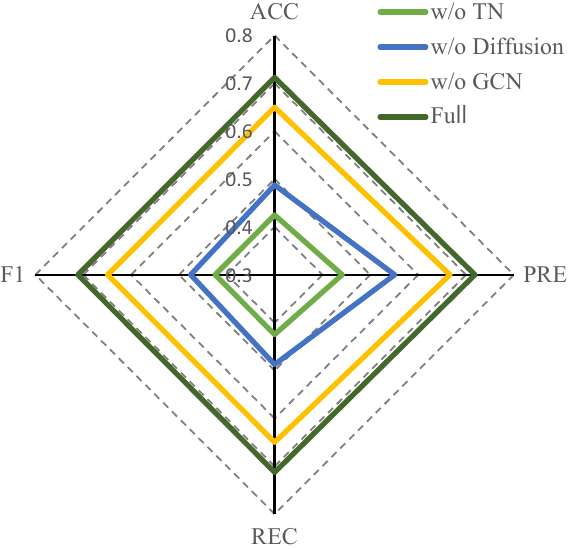}
        \medskip
    \end{minipage}
    \caption{
    Comparative analysis of classification performance across various metrics is conducted between the proposed model and PANDA, as well as the ablation study.
    }
    \label{fig:radar}
\end{figure*}

Table \ref{tab:result}  and Figure \ref{fig:radar} present a comprehensive comparison of ConnectomeDiffuser against several state-of-the-art methods for brain network construction and neurological disorder classification. Our proposed framework consistently outperformed competing approaches across both datasets, achieving superior accuracy (71.25\% for ADNI, 87.50\% for ABIDE), precision (71.84\% and 87.50\%), recall (71.25\% and 87.50\%), and F1 scores (71.09\% and 87.50\%). These results demonstrate ConnectomeDiffuser's enhanced capability for identifying distinctive biomarkers associated with neurodegenerative and neurodevelopmental disorders.

\begin{figure}[ht]
    \begin{minipage}[b]{0.49\linewidth}
        \centering
        \includegraphics[width=\linewidth]{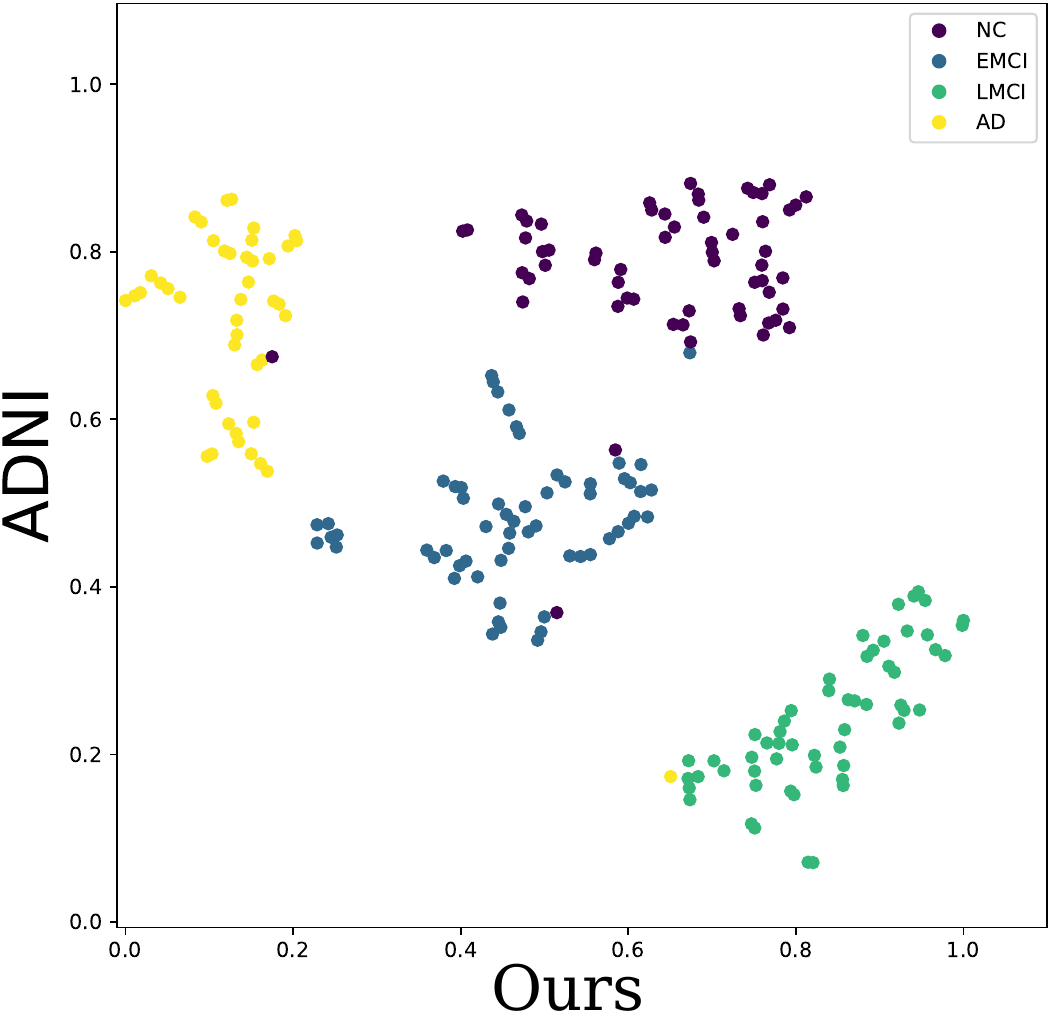}
        \smallskip
    \end{minipage}
    \begin{minipage}[b]{0.49\linewidth}
        \centering
        \includegraphics[width=\linewidth]{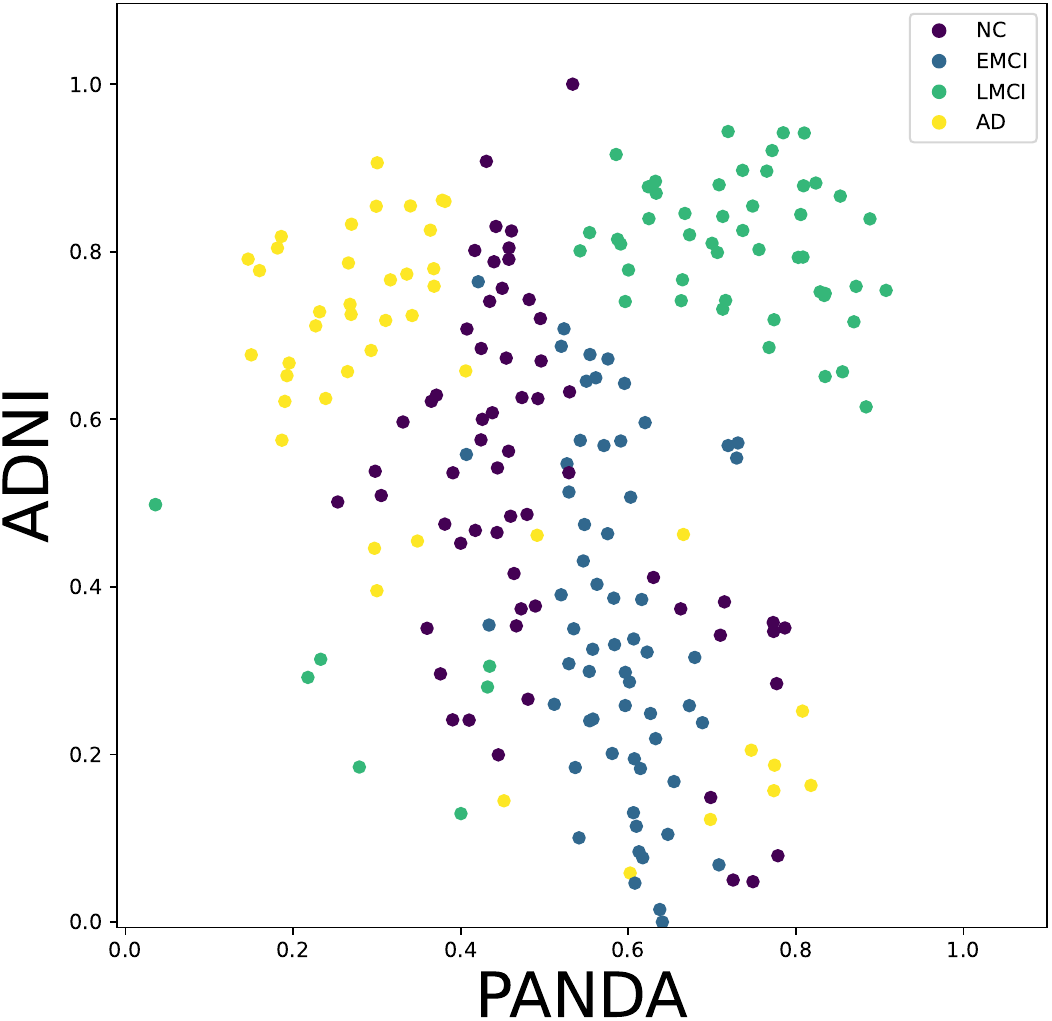}
        \smallskip
    \end{minipage}
    \begin{minipage}[b]{0.49\linewidth}
        \centering
        \includegraphics[width=\linewidth]{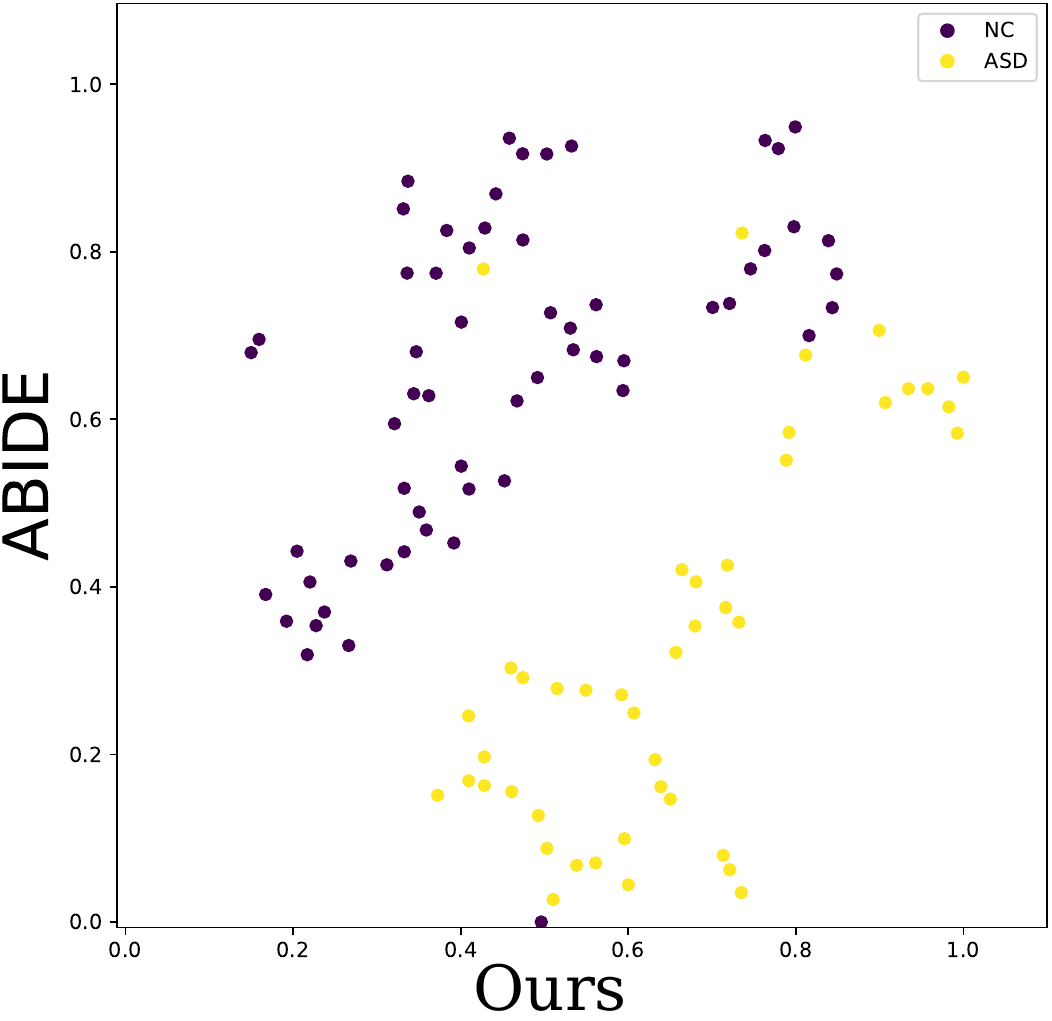}
    \end{minipage}
    \begin{minipage}[b]{0.49\linewidth}
        \centering
        \includegraphics[width=\linewidth]{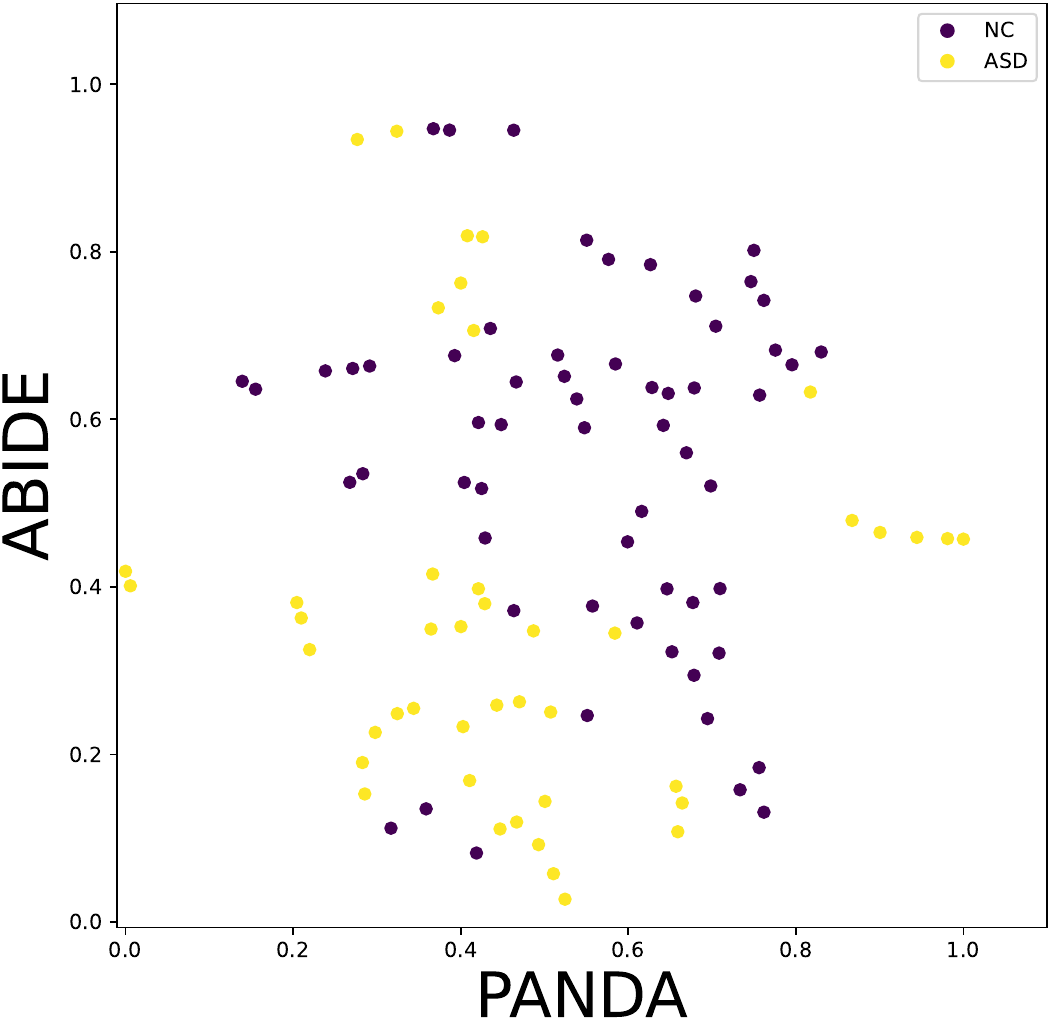}
    \end{minipage}
    \caption{
    T-SNE analysis between ConnectomeDiffuser and PANDA.
    }
    \label{fig:tsne}
\end{figure}

The performance improvement over the PANDA toolkit is particularly noteworthy, with absolute accuracy gains of 15.00\% for ADNI and 18.75\% for ABIDE. This substantial enhancement underscores the effectiveness of our end-to-end approach in capturing complex topological features and disease-specific markers directly from DTI data. The t-SNE visualization in Fig. \ref{fig:tsne} further illustrates ConnectomeDiffuser's superior discriminative capability, revealing more distinct clusters corresponding to different diagnostic categories compared to PANDA-generated networks.

The ablation study, detailed in Table \ref{tab:ab}, Table \ref{tab:ab2} and Figure \ref{fig:radar}, is conducted by the following manner:
\begin{enumerate}
    \item Remove Template Network (TN) and AAL atlas template, use normal 3D CNN instead
    \item Remove diffusion model and replace with Variational Autoencoder (VAE)
    \item Remove GCN classifier and replace with normal linear layer
    \item The impact of different numbers of GCN layer
    \item The effect of different diffusion steps
\end{enumerate}
Results demonstrate that removing the AAL template and Template Network significantly impairs the model's ability to extract meaningful topological features from brain imaging data, leading to decreased performance. The diffusion model proves superior to the VAE in generative capabilities, capturing more disease-relevant information. While the Graph Convolutional Network (GCN) classifier has a relatively modest impact on overall model performance, its removal still results in reduced classification accuracy. Regarding GCN architecture, increasing the number of layers initially improves prediction accuracy through better capture of both local and global graph structures. However, performance deteriorates beyond an optimal point due to the over-smoothing effect, where node representations become homogeneous. Prediction accuracy improves with more diffusion steps as the model better approximates the underlying data distribution, with 100 steps providing an optimal balance between computational cost and performance.

\begin{table}[ht]
\centering
\caption{Classification performance comparison of different brain network construction methods.}
\adjustbox{width=\columnwidth}{
\begin{tabular}{cccccc}
\hline
Dataset                & Method        & ACC(\%) & PRE(\%) & REC(\%) & F1(\%) \\ \hline
\multirow{6}{*}{ADNI}  & ConnectomeDiffuser & 71.25        & 71.84        &  71.25       & 71.09      \\
& SBGM         & 60.00        & 62.50        & 50.00        & 55.56       \\
& DecGAN         & 60.00        & 58.33       & 70.00        & 63.64      \\
& Brain Diffuser       & 65.00        & 60.00        & 66.67        & 63.16       \\ 
& Sang \etal         & 50.00       & 57.14        & 66.67         & 61.54       \\
                       & PANDA         & 56.25        & 56.75        & 56.25        & 55.80       \\ \hline
\multirow{6}{*}{ABIDE} & ConnectomeDiffuser & 87.50        & 87.50        & 87.50        & 87.50       \\
& SBGM         & 62.50        & 50.00        & 66.67        & 57.14       \\
& DecGAN         & 75.00        & 83.33        & 62.50        & 71.43       \\
& Brain Diffuser         & 75.00        & 66.67        & 85.71        & 75.00       \\
& Sang \etal         & 50.00        & 40.00        & 66.67        & 50.00       \\
                       & PANDA         & 68.75        & 75.00        & 66.66        & 70.59       \\ \hline
\end{tabular}
}
\label{tab:result}
\end{table}

\begin{table}[ht]
\centering
\caption{Ablation study on different modules of the model.}
\adjustbox{width=\columnwidth}{
\begin{tabular}{cccccccc}
\hline
TN & $L_{FE}$ & Diffusion & GCN & ACC(\%) & PRE(\%) & REC(\%) & F1(\%) \\ \hline
\XSolidBold             & \CheckmarkBold    & \CheckmarkBold              & \CheckmarkBold    & 42.50        & 44.13        & 42.50        & 42.43       \\
\CheckmarkBold            & \XSolidBold     & \CheckmarkBold              & \CheckmarkBold    & 63.75        & 55.26        & 63.64        & 59.15       \\
\CheckmarkBold     & \CheckmarkBold            &  \XSolidBold             & \CheckmarkBold    & 48.75        & 55.09        & 48.75        & 47.54       \\
\CheckmarkBold    & \CheckmarkBold             &  \CheckmarkBold             &  \XSolidBold   & 65.00        &  66.65       & 65.00        & 64.90       \\
\CheckmarkBold    & \CheckmarkBold             & \CheckmarkBold              & \CheckmarkBold    & 71.25        & 71.84        &  71.25       & 71.09        \\ \hline
\end{tabular}}
\label{tab:ab}
\end{table}

\begin{table}[ht]
\centering
\caption{Ablation study on number of GCN layers and diffusion steps.}
\adjustbox{width=\columnwidth}{
\begin{tabular}{cccccc}
\hline
GCN layers & Diffusion steps &  ACC(\%) & PRE(\%) & REC(\%) & F1(\%)  \\ \hline
2   & 100  & 65.00 & 59.46 & 62.86 & 61.11 \\
3   & 100  &  71.25        & 71.84        &  71.25       & 71.09  \\
4   & 100  & 68.75 & 65.79 & 67.57 & 66.67 \\
3   & 10   & 55.00 & 36.84 & 53.85 & 43.75 \\
3   & 50   & 60.00 & 47.37 & 60.00 & 52.94 \\ \hline
\end{tabular}%
}
\label{tab:ab2}
\end{table}

\begin{figure*}[ht]
    \begin{minipage}[b]{0.32\linewidth}
        \centering
        \includegraphics[width=\linewidth]{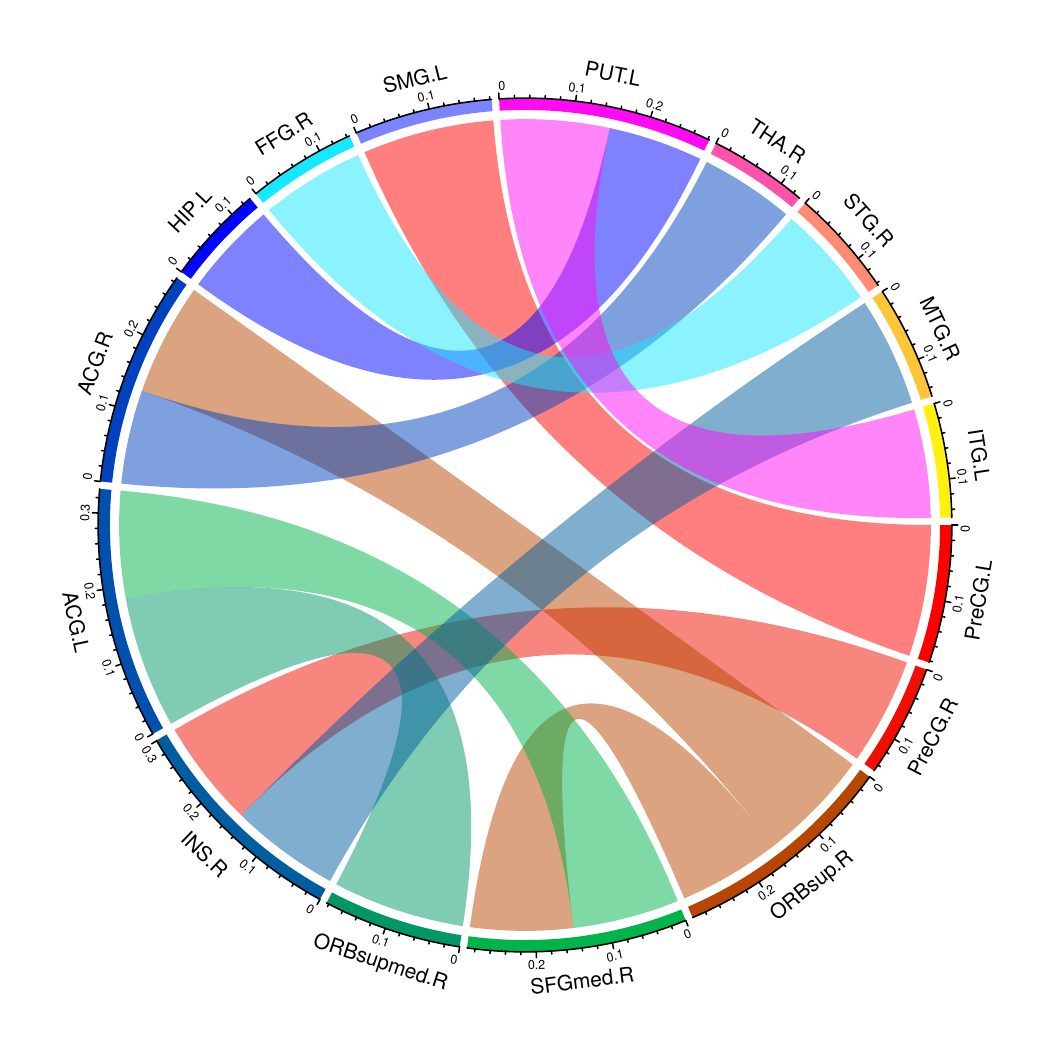}
        \medskip
    \end{minipage}
    \begin{minipage}[b]{0.32\linewidth}
        \centering
        \includegraphics[width=\linewidth]{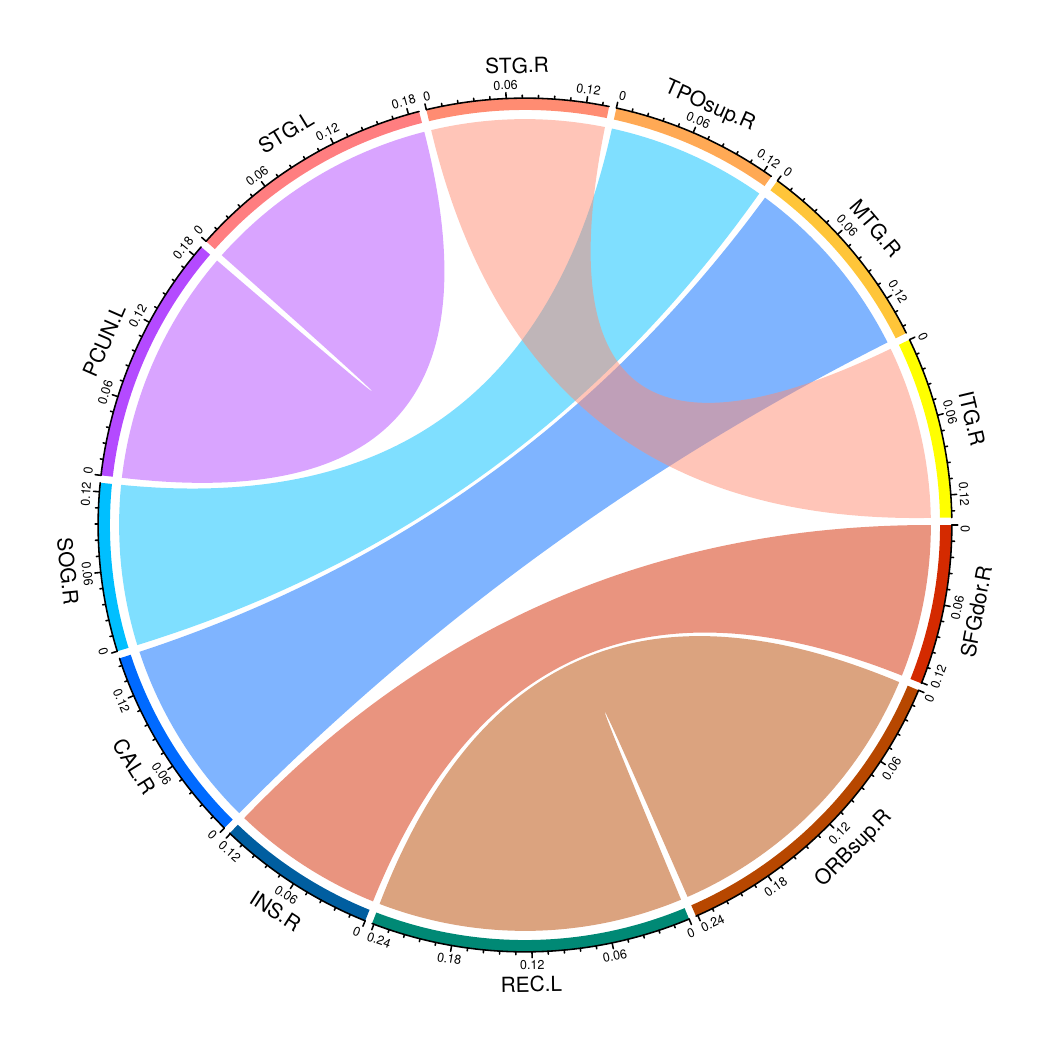}
        \medskip
    \end{minipage}
    \begin{minipage}[b]{0.32\linewidth}
        \centering
        \includegraphics[width=\linewidth]{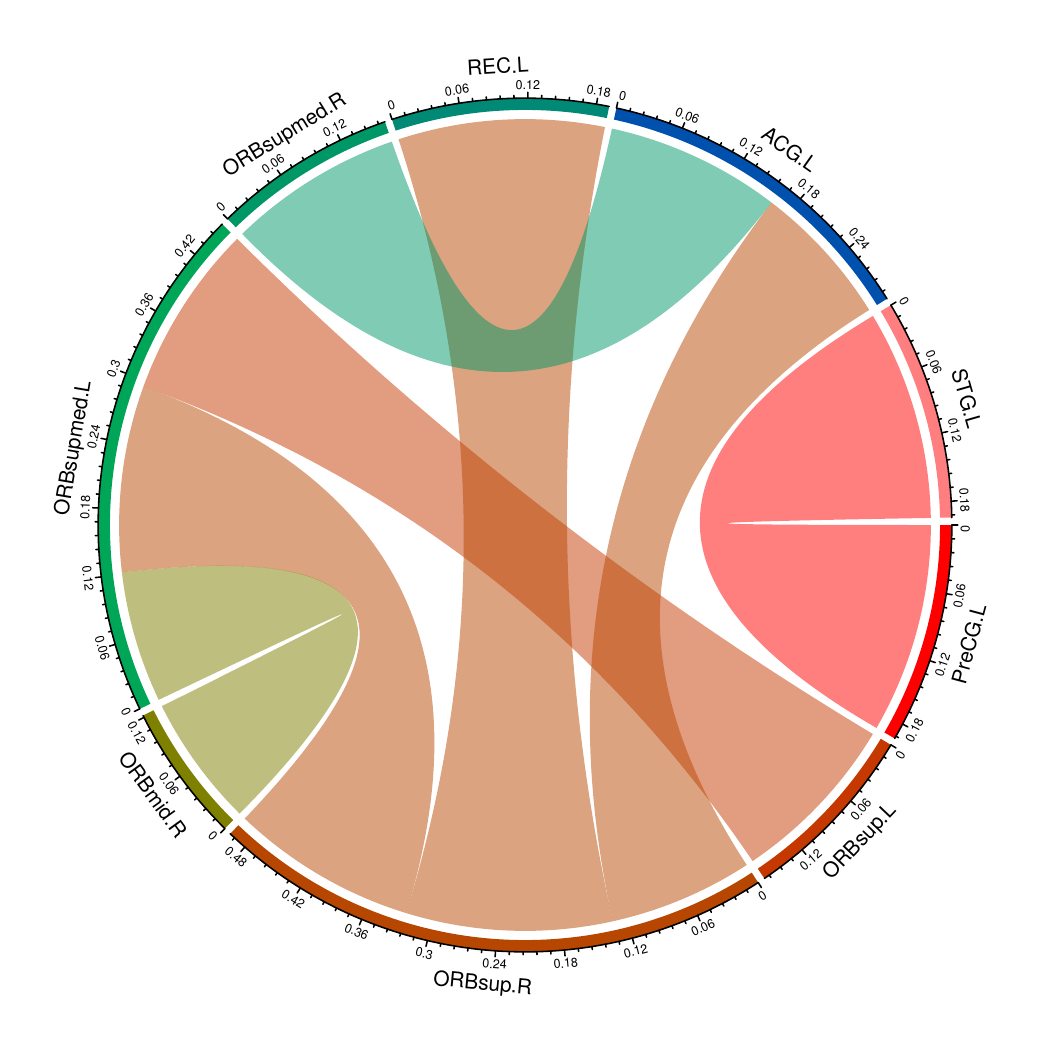}
        \medskip
    \end{minipage}
    \begin{minipage}[b]{0.32\linewidth}
        \centering
        \includegraphics[width=\linewidth]{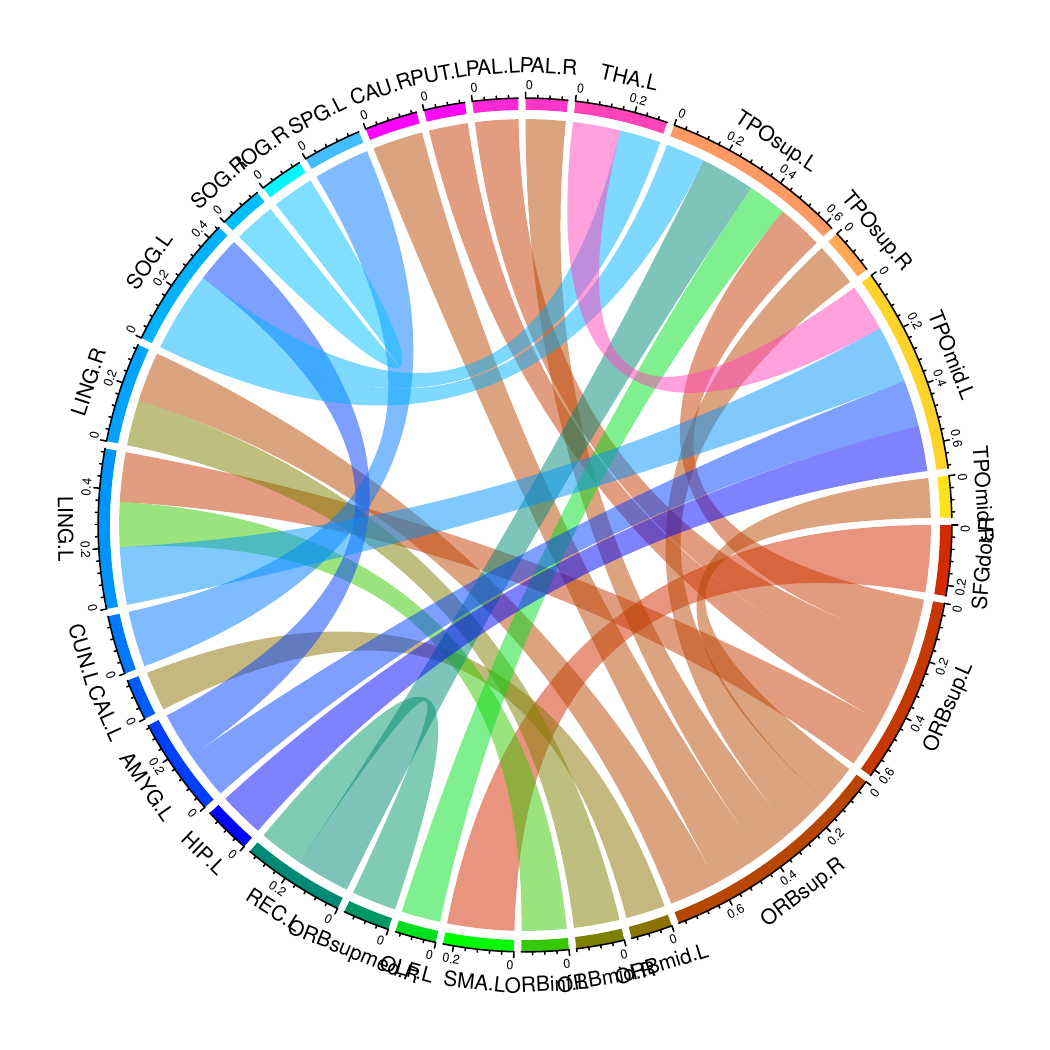}
        \centerline{NC vs. EMCI}\medskip
    \end{minipage}
    \begin{minipage}[b]{0.32\linewidth}
        \centering
        \includegraphics[width=\linewidth]{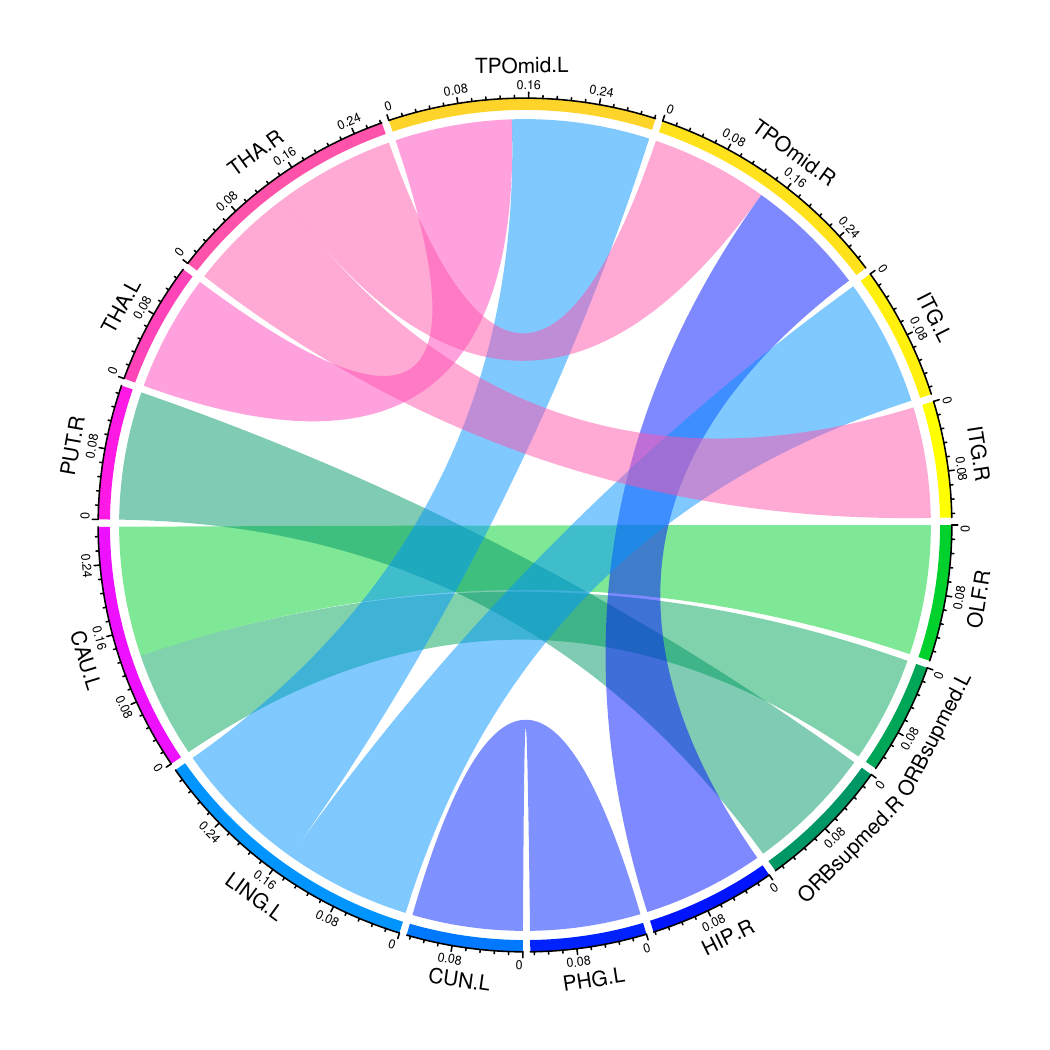}
        \centerline{EMCI vs. LMCI}\medskip
    \end{minipage}
    \begin{minipage}[b]{0.32\linewidth}
        \centering
        \includegraphics[width=\linewidth]{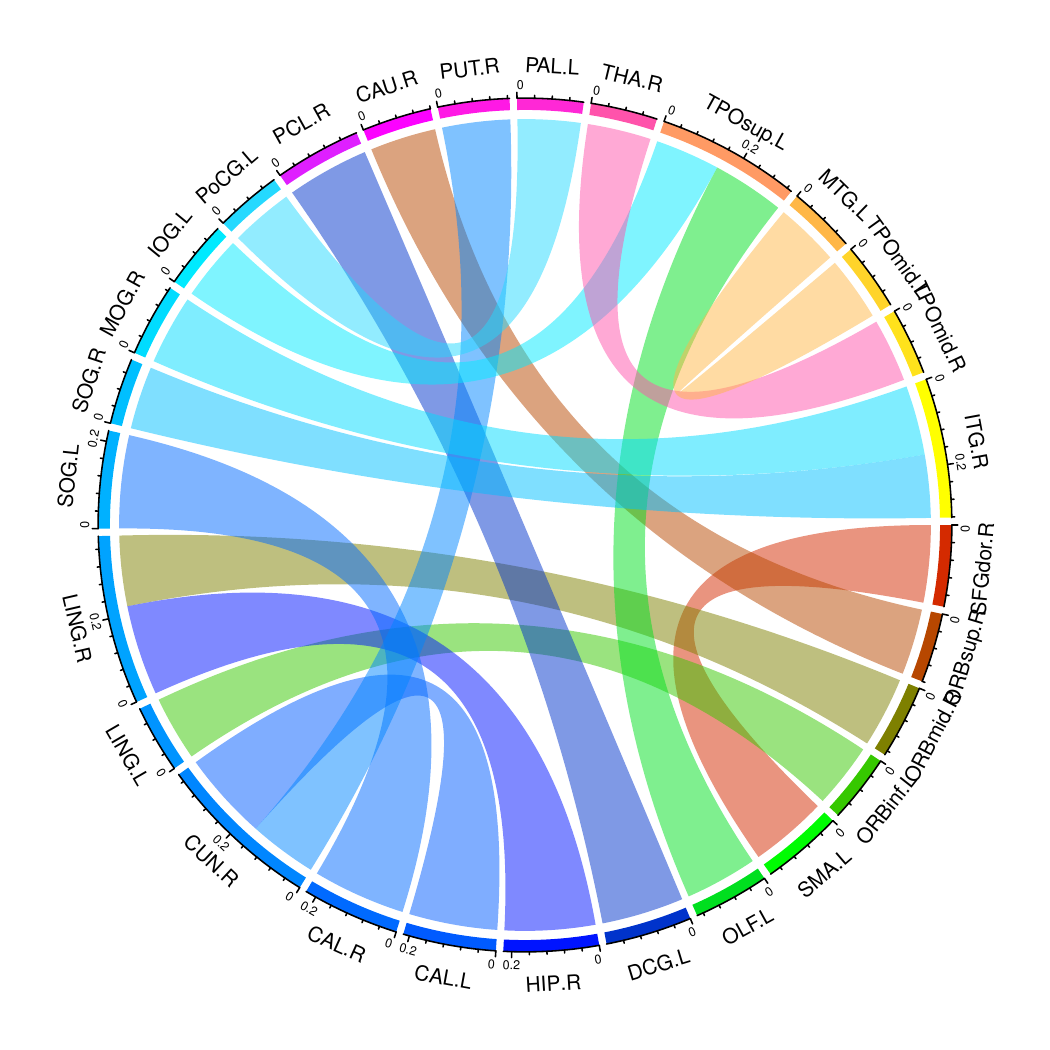}
        \centerline{LMCI vs. AD}\medskip
    \end{minipage}
    \caption{
    Chord diagram of the connectivity difference in the ADNI dataset. The top row represents the increased connectivity, from left to right: EMCI to NC, LMCI to NC and AD to NC. The bottom row represents the decreased connectivity, from left to right: EMCI to NC, LMCI to NC and AD to NC.
    }
    \label{fig:chord1}
\end{figure*}

\begin{figure*}[ht]
    \begin{minipage}[b]{0.32\linewidth}
        \centering
        \includegraphics[width=\linewidth]{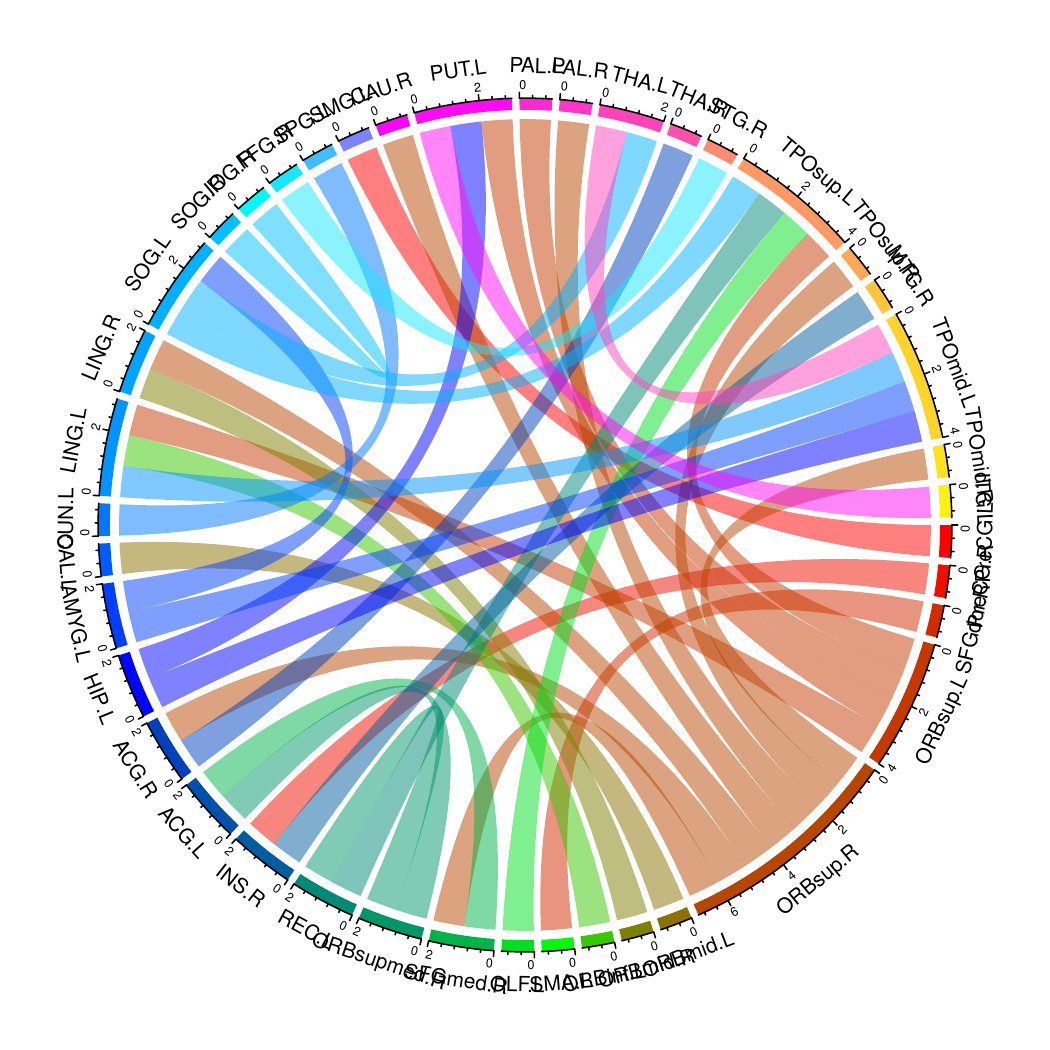}
        \centerline{NC vs. EMCI}\medskip
    \end{minipage}
    \begin{minipage}[b]{0.32\linewidth}
        \centering
        \includegraphics[width=\linewidth]{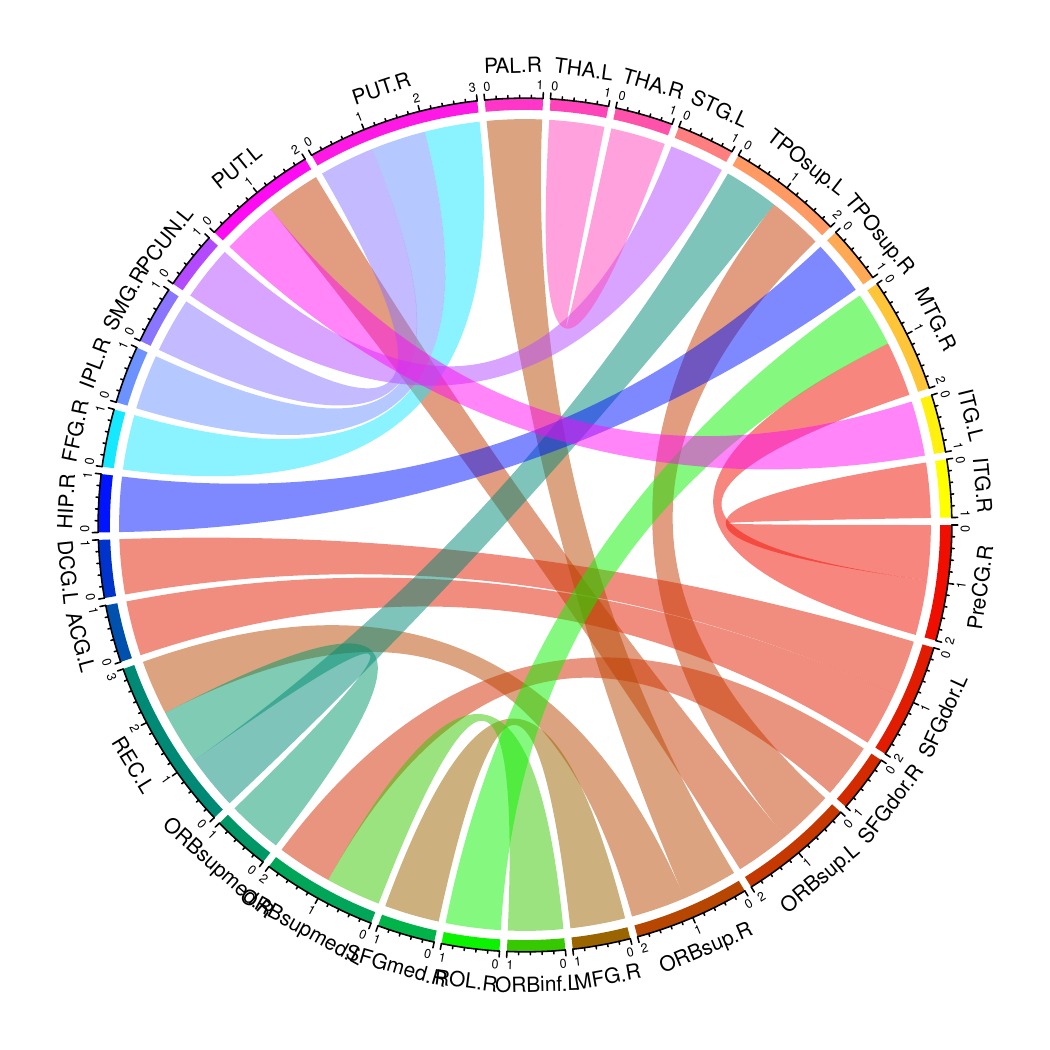}
        \centerline{EMCI vs. LMCI}\medskip
    \end{minipage}
    \begin{minipage}[b]{0.32\linewidth}
        \centering
        \includegraphics[width=\linewidth]{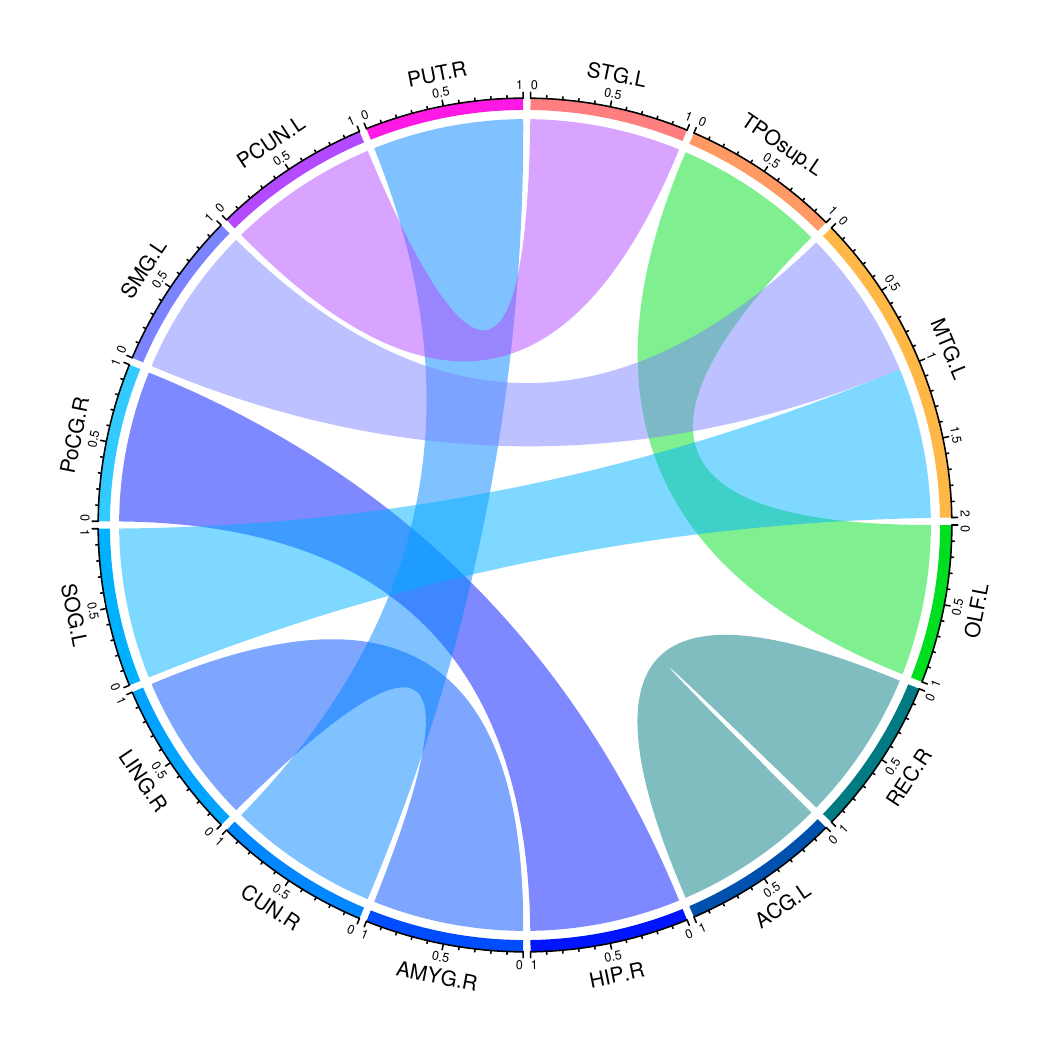}
        \centerline{LMCI vs. AD}\medskip
    \end{minipage}
    \caption{
    Chord diagram of the alterations in neural connectivity in the ADNI dataset. The left side represents the connectivity changes between NC and EMCI. The middle section displays the connectivity changes between EMCI and LMCI. The right side illustrates the connectivity changes between LMCI and AD.
    }
    \label{fig:chord2}
\end{figure*}

\subsection{Structural Brain Network Connectivity Analysis of Alzhemier's Disease}
We leveraged ConnectomeDiffuser to investigate the progressive alterations in brain connectivity patterns across the Alzheimer's disease stages. For each subject in the ADNI test set, we generated structural brain networks and subsequently employed two-sample t-tests to identify statistically significant connectivity differences (p $<$ 0.05) between diagnostic groups. Connections failing to meet this significance threshold were considered disease-irrelevant and subsequently nullified.

To elucidate the subtle connectivity alterations associated with disease progression, we constructed average brain networks for each diagnostic category and computed inter-group differences. These differential patterns were then correlated with the abnormal regions identified via statistical testing, revealing distinct patterns of connectivity enhancement and deterioration across disease stages, as visualized in Figure \ref{fig:chord1}.

Our analysis revealed a striking imbalance between increased and decreased connectivity as AD progresses, with connectivity reductions becoming increasingly predominant at more advanced disease stages. This pattern aligns with the neuropathological trajectory of AD, characterized by progressive neuronal loss and brain tissue atrophy that systematically disrupts structural connections between brain regions. Figure \ref{fig:chord2} illustrates that the magnitude of neural connectivity alterations diminishes as the disease advances from EMCI to LMCI to AD, suggesting that major connectivity disruptions occur early in the disease process, with fewer additional changes as neurodegeneration progresses to advanced stages.

Interestingly, we observed selective increases in connectivity among specific brain regions despite the overall trend toward reduced connectivity. This phenomenon likely represents compensatory mechanisms whereby the brain recruits alternative neural pathways to maintain cognitive function despite accumulating neuropathology, particularly in earlier disease stages.

\begin{figure}[ht]
    \begin{minipage}[b]{0.49\linewidth}
        \centering
        \includegraphics[width=\linewidth]{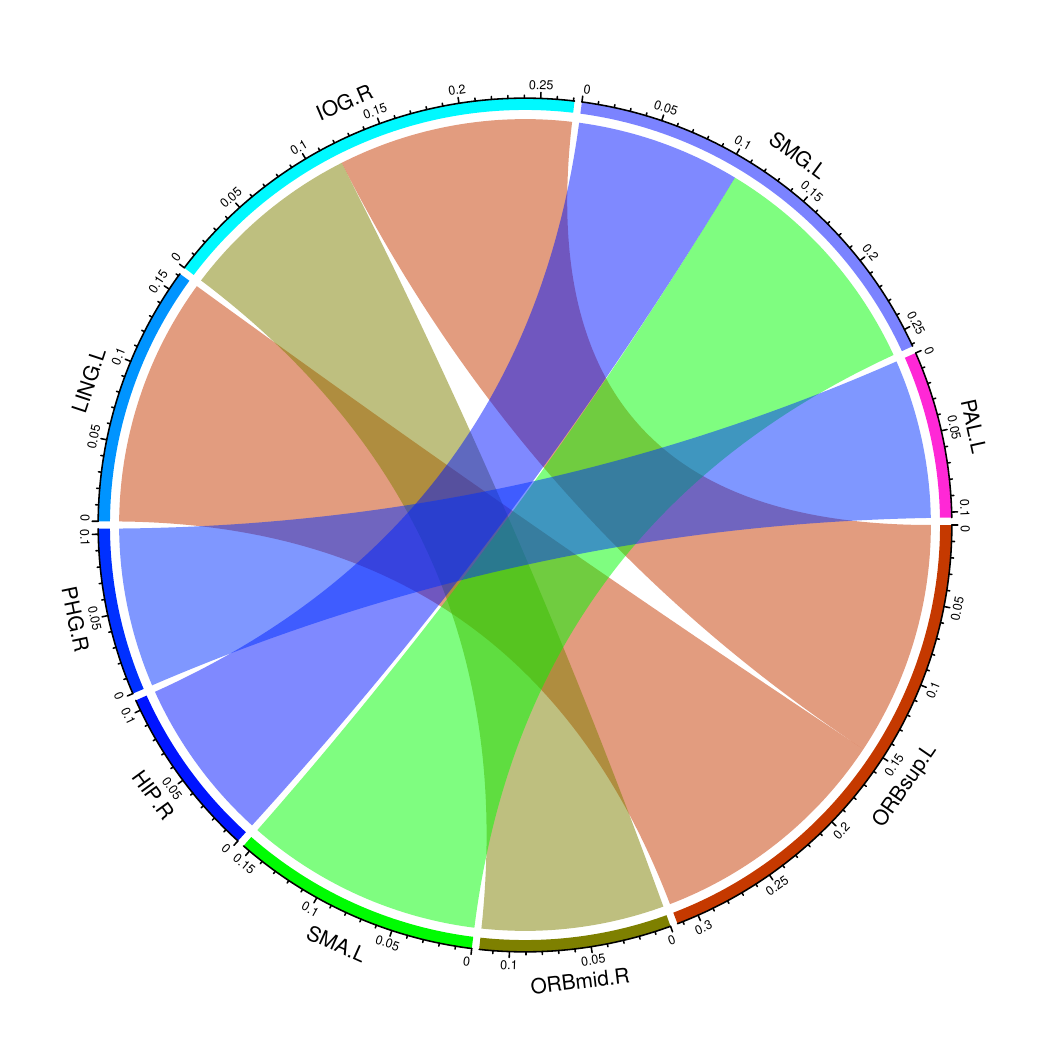}
        \centerline{Increased connectivity}\medskip
    \end{minipage}
    \begin{minipage}[b]{0.49\linewidth}
        \centering
        \includegraphics[width=\linewidth]{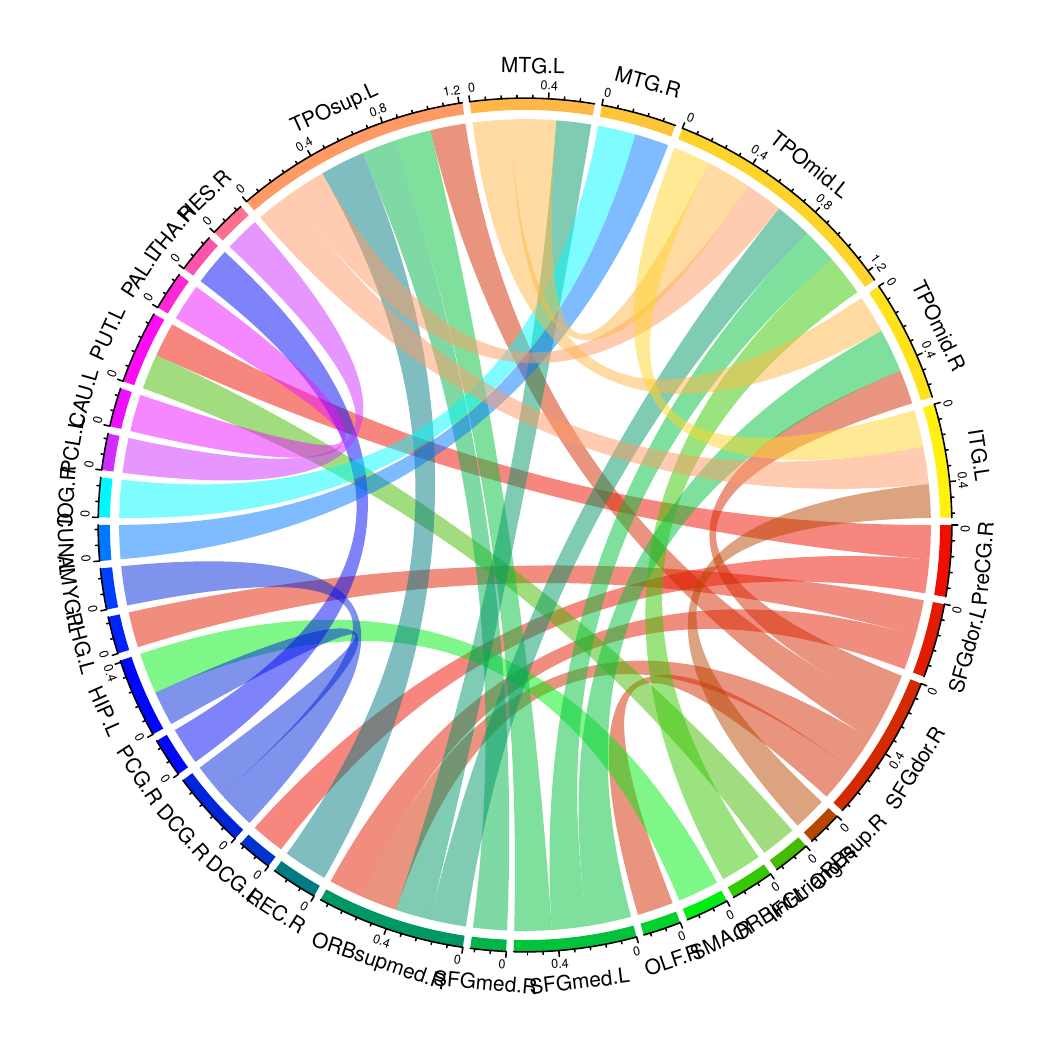}
        \centerline{Decreased connectivity}\medskip
    \end{minipage}
    \caption{
    Chord diagram of the connectivity difference in the ABIDE dataset. The left side of the diagram illustrates an increase in connectivity from NC to ASD, while the right side demonstrates a decrease in connectivity from NC to ASD.
    }
    \label{fig:chord3}
\end{figure}

\subsection{Structural Brain Network Connectivity Analysis of Autism Spectrum Disorder}
We extended our connectivity analysis methodology to the ABIDE dataset, examining structural brain network differences between individuals with ASD and matched neurotypical controls. Following network construction using ConnectomeDiffuser, we applied the same statistical method to identify significant connectivity alterations between groups.

In contrast to the predominantly decreasing connectivity observed in AD progression, ASD demonstrated a more balanced pattern of connectivity changes, with comparable numbers of significantly increased and decreased connections (Figure \ref{fig:chord3}). This distinct pattern suggests that ASD involves specific reconfiguration of structural connectivity rather than wholesale connectivity reduction, consistent with contemporary theories characterizing ASD as a disorder of brain connectivity.

The more balanced connectivity alterations in ASD may also reflect neurodevelopmental adaptations, particularly given the adolescent age range of the ABIDE cohort. During this period of heightened neuroplasticity, compensatory mechanisms may establish long-range connections to counterbalance local connectivity disruptions characteristic of ASD, potentially facilitating the development of alternative cognitive strategies.

\subsection{Advanced Analysis of Structural Brain Network Construction Efficacy}
Beyond classification performance, we evaluated ConnectomeDiffuser's practical utility by assessing its computational efficiency and output reproducibility compared to the conventional PANDA toolkit. Table \ref{tab:vspad} presents a comparative analysis focusing on processing time and result consistency.

The PANDA toolkit requires extensive manual parameter calibration and complex computational procedures for structural brain network construction. The reported processing time (900 seconds per subject) encompasses both computational operations and necessary manual interventions, including parameter tuning and data preparation. This substantial time investment varies considerably (900-1500 seconds) depending on operator expertise, introducing potential workflow inefficiencies.

In contrast, ConnectomeDiffuser streamlines this process through its end-to-end design, requiring merely 2.5 seconds per subject for inference following the one-time offline training phase. This represents a 360-fold reduction in processing time, dramatically enhancing throughput for large-scale neuroimaging studies and clinical applications.

Importantly, ConnectomeDiffuser's data-driven approach eliminates the need for subjective parameter tuning, mitigating human bias and enhancing result objectivity. We quantitatively assessed result consistency using the Intra-Class Correlation (ICC) coefficient \cite{hu2023phipipe,zong2024new}, which measures the reliability and reproducibility of measurements. ConnectomeDiffuser demonstrated superior ICC values (0.76 for ADNI, 0.68 for ABIDE) compared to PANDA (0.73 and 0.64, respectively), confirming its enhanced reproducibility. This improvement is particularly valuable in clinical settings, where consistent and reliable results are paramount for accurate diagnosis and treatment planning.

\begin{table}[ht]
\centering
\caption{Advanced analysis of PANDA and ConnectomeDiffuser.}
\begin{tabular}{ccccc}
\hline
Dataset                & Method        & Time(s) & Intra-Class Correlation \\ \hline
\multirow{2}{*}{ADNI}  & PANDA         & 900     &  0.73       \\
                       & ConnectomeDiffuser & 2.5     & 0.76        \\
\multirow{2}{*}{ABIDE} & PANDA         & 900     & 0.64       \\
                       & ConnectomeDiffuser & 2.5     & 0.68        \\ \hline
\end{tabular}
\label{tab:vspad}
\end{table}

\subsection{Limitations}
Despite its compelling performance, ConnectomeDiffuser has several limitations worth acknowledging. First, the framework relies on the AAL atlas as a standardized template, which may not optimally capture individual anatomical variations. While this approach facilitates cross-subject comparability, it potentially obscures subject-specific anatomical nuances that could be clinically relevant.

Second, the current implementation focuses exclusively on structural brain networks derived from DTI. This single modality approach does not capitalize on the complementary information available from functional connectivity, which has demonstrated significant value in characterizing neurological disorders. Integrating functional connectivity data could potentially enhance diagnostic precision and provide more comprehensive insights into disease mechanisms.

These limitations present promising directions for future research, including the incorporation of subject-specific anatomical templates and multimodal integration strategies to further enhance the clinical utility of ConnectomeDiffuser.

\section{Conclusion}
In this study, we introduce ConnectomeDiffuser, an innovative end-to-end model for the construction of individualized structural brain networks derived from DTI. Utilizing the AAL atlas as a structural guide, we harness a Template Network to effectively capture topological features. ConnectomeDiffuser then synthesizes these features, along with disease-specific labels from a classifier, to produce the brain networks. Our experimental evaluations, conducted on the ADNI and ABIDE datasets, reveal that ConnectomeDiffuser surpasses existing methods in terms of its analytical capabilities and predictive accuracy regarding the progression of various brain disorders. A significant finding of our model is the tendency for a reduction in brain connectivity as diseases progress. This insight has the potential to contribute to a deeper understanding of the pathophysiological mechanisms of brain disorders and may facilitate the identification of early biomarkers critical for targeted therapeutic interventions.

\bibliographystyle{IEEEtranDOI}
\bibliography{ref}

\end{document}